\documentclass[lettersize,journal]{IEEEtran}

\usepackage{amsthm}

\newtheorem*{prob_state}{\textbf{Problem Statement}}

\usepackage{graphicx}
\usepackage[caption=false,font=normalsize,labelfont=sf,textfont=sf]{subfig}
\usepackage{amsmath,amsfonts}
\usepackage{amssymb}
\usepackage[ruled, vlined, linesnumbered]{algorithm2e}
\usepackage[hidelinks]{hyperref}
\usepackage{cite}
\usepackage{makecell}
\usepackage[numbers]{natbib}
\usepackage{array}
\usepackage{tikz}
\usetikzlibrary{trees}
\usepackage{tabularx}
\usepackage{threeparttable}
\usepackage{booktabs}
\usepackage{multirow}
\usepackage[capitalize,noabbrev]{cleveref}
\crefname{equation}{Eq.}{Eqs.}
\usepackage{enumitem}
\usepackage{xcolor}

\usepackage{dblfloatfix}

\usepackage{url}

\definecolor{DeepPink}{HTML}{FF1493}
\definecolor{Orchid}{HTML}{DA70D6}
\definecolor{Magenta}{HTML}{FF00FF}
\definecolor{Fuchsia}{HTML}{FF00FF}
\definecolor{LavenderPink}{HTML}{FFB6C1}

\begin{document}

\title{SCU: An Efficient Machine Unlearning Scheme for Deep Learning Enabled Semantic Communications}

\author{Weiqi~Wang,~\IEEEmembership{Member,~IEEE},
	Zhiyi~Tian,~\IEEEmembership{Member,~IEEE},
	Chenhan~Zhang,~\IEEEmembership{Member,~IEEE},
	and Shui~Yu,~\IEEEmembership{Fellow,~IEEE}

\thanks{This paper was supported in part by Australia ARC LP220100453, ARC DP200101374, and ARC DP240100955. \it{(Corresponding author: Zhiyi Tian.)}}
\IEEEcompsocitemizethanks{\IEEEcompsocthanksitem W. Wang, Z. Tian and S. Yu are with the School of Computer Science, University of Technology Sydney, Australia. C. Zhang is with the Postdoctoral Research Fellow, Macquarie University, Australia \protect\\
	E-mail: {Weiqi.Wang-2@student.uts.edu.au, chzhang@ieee.org, \{zhiyi.tian-1, shui.yu\}@uts.edu.au}
}
}

\markboth{This paper is accepted by IEEE Transactions on Information Forensics and Security}%
{Shell \MakeLowercase{\textit{et al.}}: A Sample Article Using IEEEtran.cls for IEEE Journals}


\maketitle

\begin{abstract}

Deep learning (DL) enabled semantic communications leverage DL to train encoders and decoders (codecs) to extract and recover semantic information. However, most semantic training datasets contain personal private information. Such concerns call for enormous requirements for specified data erasure from semantic codecs when previous users hope to move their data from the semantic system. {Existing machine unlearning solutions remove data contribution from trained models, yet usually in supervised sole model scenarios. These methods are infeasible in semantic communications that often need to jointly train unsupervised encoders and decoders.} In this paper, we investigate the unlearning problem in DL-enabled semantic communications and propose a semantic communication unlearning (SCU) scheme to tackle the problem. {SCU includes two key components. Firstly,} we customize the joint unlearning method for semantic codecs, including the encoder and decoder, by minimizing mutual information between the learned semantic representation and the erased samples. {Secondly,} to compensate for semantic model utility degradation caused by unlearning, we propose a contrastive compensation method, which considers the erased data as the negative samples and the remaining data as the positive samples to retrain the unlearned semantic models contrastively. Theoretical analysis and extensive experimental results on three representative datasets demonstrate the effectiveness and efficiency of our proposed methods.

\end{abstract}

\begin{IEEEkeywords}
	Semantic communication, machine unlearning.
\end{IEEEkeywords}

\section{Introduction}
 
\begin{figure}
	\centering
	\vspace{0mm}
	\includegraphics[width=0.85\linewidth]{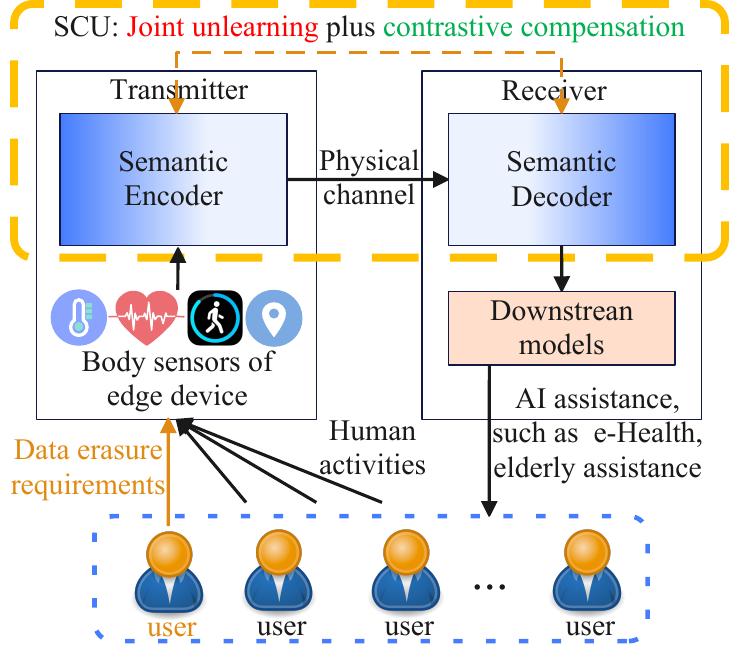}
	\caption{Motivation and unlearning requirements in DL-enabled semantic communication systems. In healthcare scenarios, the server needs to collect users' sensitive information, such as blood pressure, heart rate, etc, for SC model training. Users also benefit from the downstream applications when the SC models are well-trained. However, there are also a huge amount of unlearning requirements from users who hope to remove their private information from trained SC models to protect their privacy.}
	\label{fig_unlearninghighlevel}
	\vspace{-2mm}
\end{figure}

\IEEEPARstart{S}{emantic} communication has attracted significant attention recently. It is regarded as a significant advancement beyond the Shannon paradigm, as semantic communication focuses on transmitting the underlying semantic information from the source, rather than ensuring the accurate reception of each individual symbol or bit irrespective of its meaning \cite{lan2021semantic,qin2021semantic}. With the burgeoning advancement of deep learning (DL), researchers found that employing DL models as the encoder and decoder greatly improves semantic transmission efficiency and reliability \cite{zhang2022deep,zhang2023predictive}, called DL-enabled semantic communications. However, to train these DL semantic encoders and decoders, transmitters and receivers must first collect the training datasets from huge amounts of human activities from users \cite{lan2021semantic}, which contain rich personal privacy information. Once users want to quit the semantic communication service, it requires the server to remove their sensitive information from trained semantic models, ensuring users' right to be forgotten \cite{cao2015towards,voigt2017eu}. For example, in \Cref{fig_unlearninghighlevel}, in semantic communications for healthcare scenarios \cite{ning2020mobile,liu2023survey,lan2021semantic}, people use body sensors to collect information, such as blood pressure, heart rate, walking steps, etc., for the semantic models training. These sensitive information must be unlearned from semantic models if users want to quit the service. 




{However, there are three key challenges when unlearning semantic communication models. First, semantic communications are usually implemented by jointly trained DL encoder and decoder \cite{zhang2023predictive,qin2021semantic}. Most existing machine unlearning methods mainly focus on unlearning single model structure scenarios \cite{nguyen2020variational,sekhari2021remember}. It is challenging to maintain semantic knowledge consistency between jointly trained encoders and decoders during and after unlearning. Second, unlike conventional unlearning targeting supervised learning scenarios \cite{bourtoule2021machine}, unlearning for DL-enabled semantic models should consider the unsupervised or self-supervised learning tasks. This consideration is crucial because, across various applications of semantic communication, unsupervised learning techniques are frequently utilized for tasks such as feature extraction, dimensionality reduction, and data reconstruction \cite{huang2022toward,hu2023robust}. Third, it lacks an effective evaluation for the semantic communication unlearning methods. Existing evaluation methods \cite{hu2022membership} for the unlearning supervised models are infeasible to the semantic models with an unsupervised recovery task \cite{hu2023robust,huang2022toward}. 
}




\vspace{2mm}
\noindent
\textbf{Our work:} In this paper, we aim to overcome the above challenges and propose a semantic communication unlearning (SCU) method, as shown in \Cref{fig_unlearninghighlevel}. Our SCU unlearns semantic models by removing the information of specified samples from the learned semantic representation between encoders and decoders. {It includes two main components,  joint unlearning and contrastive compensation.} To make our method easy to understand, we choose to introduce our SCU method following the variational information bottleneck (VIB) framework \cite{alemi2016deep}, and all the original models before unlearning are trained using the VIB method. 
VIB and variational autoencoder (VAE) \cite{hu2023robust} are common unsupervised DL methods that are often used in semantic communications, and these two mainstream unsupervised methods are proved to have the same expansion mathematic form \cite{alemi2016deep}. 





{In SCU, we first} design a joint unlearning method, simultaneously implementing data erasure from the semantic encoder and decoder. We unlearn the semantic models through the optimization of mutual information. Specifically, the joint unlearning method in SCU unlearns the encoder and decoder by minimizing mutual information between the learned semantic representation and the specified unlearning samples. Although joint unlearning achieves a slightly better or similar performance compared with existing unlearning methods in unsupervised semantic communications, it also causes model utility degradation, which occurs in most approximate unlearning methods. {Then,} we propose a contrastive compensation method to compensate for the utility degradation, ensuring that the semantic knowledge learned on the remaining dataset is consistent in the encoder and decoder. 
The contrastive compensation considers the erased data as the negative samples and the remaining data as the positive samples to help retrain the unlearned semantic models to maintain the performance. 
Through this approach, we ensure that the unlearned semantic models forget the erased samples, yet they still preserve the semantic knowledge consistency learned before.

{To effectively evaluate unlearning of semantic communications, two evaluation metrics are proposed: the reconstruction mean squared error (MSE) comparison and the downstream backdoor detection.} Specifically, we mixed the prepared backdoored samples into the full training dataset to train the semantic codecs. Then, we train a downstream classification model to identify the backdoor information of the decoder's outputs. The goal of the unlearning methods is to erase the impact of the backdoored data points from trained semantic models. {The effectiveness of unlearning can be measured via the backdoor detection accuracy of the downstream classification model. A low backdoor accuracy of the downstream model means a high unlearning effect.} We also compare the MSE between the reconstructed and original samples to check whether the recovered data contains the backdoor. Comprehensive experimental results show that SCU greatly improves effectiveness and efficiency over the baselines by directly applying state-of-the-art unlearning methods in semantic communications.


Our contributions are summarized as follows.
\begin{itemize}
	\item We formally define the unlearning problem in semantic communication, {focusing on simultaneously unlearning the semantic encoder and decoder for the trained unsupervised semantic codecs.}
	\item We propose an effective method, SCU, to tackle the semantic unlearning problem, where we customize the joint unlearning for data erasure of both the semantic encoder and decoder simultaneously. Moreover, we propose a contrastive compensation method to compensate for the model utility degradation caused by unlearning. 
	\item We conduct a comprehensive analysis and customize two evaluation metrics for semantic communication unlearning methods. Extensive experimental results demonstrate the proposed SCU's well-performed unlearning effectiveness and efficiency. 
	\item {The source code and the artifact of SCU are released at \url{https://github.com/wwq5-code/SCU}, which creates a new tool for implementing machine unlearning for unsupervised DL-enabled semantic communications and sheds light on the design of unsupervised DL scenarios.}
\end{itemize}

The rest of the paper is structured as follows. Related work is discussed in \Cref{rw}. Background knowledge on training semantic communication models and the problem definition of semantic communication unlearning are covered in \Cref{pr}. The proposed solution, SCU, is detailed in \Cref{sec_SCU}. In \Cref{theoretical}, we analyze the effectiveness of unlearning as well as its time and space complexity. Experimental results and comparisons with state-of-the-art methods are presented in \Cref{ex}. The paper concludes with a summary in \Cref{summary}.


\section{Related Work}\label{rw}

\subsection{Semantic Communication} 
With the advancement of DL, numerous DL-enabled semantic communication systems have been introduced. In such semantic communication systems, DL models are utilized to autonomously learn the extraction and transmission of semantic knowledge, tailored to the data distribution, specific tasks, and the status of the communication channel \cite{lan2021semantic,qin2021semantic,zhang2022deep,zhang2023predictive}. 
There are two main kinds of semantic communication systems: task-oriented systems and full data reconstruction systems. The task-oriented semantic communications train the decoder for specified tasks, such as face identification and speech recognition \cite{xie2022task,xie2021deep,xie2021task,zhang2023toward}. In contrast, the full data reconstruction systems train the decoder to recover the source data \cite{hu2023robust,zhang2022deep,weng2021semantic,zhang2023predictive}, which can be used in downstream tasks efficiently.
Therefore, task-oriented semantic communications usually employ supervised methods to train the employed DL models, while full data reconstruction systems train models in an unsupervised way. 
Most existing unlearning methods focus on supervised scenarios \cite{nguyen2020variational,sekhari2021remember}, which provide possible solutions in task-oriented semantic communications. Full data reconstruction semantic communications \cite{hu2023robust,huang2022toward,nan2023physical} are also an important component, but there is a lack of sufficient studies about unsupervised unlearning to support full data reconstruction semantic communications. 


\subsection{Machine Unlearning}
Existing machine unlearning methods give insightful inspiration to design unlearning methods for DL-enabled semantic communications, as most of them employ DL models to extract and recover the semantic information \cite{zhang2022deep,zhang2023predictive,weng2021semantic}. There are two rough categories of machine unlearning methods: fast retraining and approximate unlearning, according to the methodology and target problems.





The fast retraining methods are extended based on the naive retraining method to reduce the computational costs \cite{cao2015towards,wu2020deltagrad}. They split the whole training dataset into multi-subsets and divided the learning process by training multiple sub-models based on the split sub-datasets. So that when the unlearning requests come, they only need to retrain the corresponding sub-model that is trained based on the sub-dataset with the erased samples, which significantly reduces the unlearning computation cost \cite{cao2015towards,bourtoule2021machine,yanarcane2022unlearning}. However, these methods are impractical for frequent unlearning requests or when the data to be erased is distributed across multiple subsets.


Approximate unlearning seeks to create a model that closely resembles one retrained solely on the remaining dataset \cite{guo2019certified,nguyen2020variational,kim2022efficient}. Two mainstream approximate unlearning methods are designed based on the Hessian matrix \cite{guo2019certified,sekhari2021remember} and variation Bayes \cite{nguyen2020variational,fu2022knowledge,nguyen2022markov}. They learn a differential similar model and variational approximate posterior, respectively. These methods are easy to cause huge model utility degradation, named catastrophic unlearning. To minimize the decrease in utility, a fixed threshold is typically employed to regulate the extent of unlearning \cite{nguyen2020variational,guo2019certified}. {Moreover, \citeauthor{fan2023salun} and \citeauthor{liu2024model} \cite{fan2023salun,liu2024model} find that considering model sparsity based on existing unlearning methods can effectively enhance the unlearning effect while not significantly decreasing model utility. Other methods, such as \cite{kurmanji2024towards,chundawat2023can,wang2023machine}, considered both the unlearning effect and model utility during the unlearning process and proposed corresponding methods to achieve the optimal balance.} 

However, all these methods are mainly designed for solely trained ML models in supervised scenarios. 
While in DL-enabled semantic communications, they usually jointly train two DL models, and there are huge amounts of unsupervised recovering tasks \cite{hu2023robust,zhang2022deep}. Unlearning for these kinds of jointly-trained DL-enabled semantic communication and other unsupervised models still requires further exploration.

\section{Preliminary and Problem Statement} \label{pr}


\subsection{Notations}
The primary notations used in this paper are summarized first, as outlined in \Cref{notation}.
Since we focus on unsupervised DL-enabled semantic communications, the full training dataset $D$ only contains input $X$. The erased dataset and the remaining dataset are $D_e$ and $D_r$. We will employ unsupervised variational information bottleneck methods to learn the semantic representation $Z$, and the corresponding persample from $X$ and $Z$ are x and z. The trained semantic model is $\mathcal{M}(f_{\theta}, g_{\theta})$, which includes an encoder $f_{\theta}$ and decoder $g_{\theta}$. We denote the unlearned model as $\mathcal{M}_u$. In our method, we will set the fixed encoder and decoder, which are denoted as $f_{\theta}^{fix}$ and $g_{\theta}^{fix}$. Our joint unlearning loss is $\mathcal{L}_{JU}$, and contrastive compensation loss is $\mathcal{L}_{CC}$. During unlearning training, we will set the corresponding weights $\alpha_1$ and $\alpha_2$ for the $\mathcal{L}_{JU}$ and $\mathcal{L}_{CC}$.


\begin{table}[t]
	\scriptsize
	\centering  
	\caption{Basic Notations}
	\label{notation}
	\resizebox{\linewidth}{!}{ 
		\begin{tabular}{c || c }  
			\toprule[0.12em]
			Notations &Descriptions \\ 
			\midrule  
			$D=X$ & \makecell[c]{the full training dataset $D$ only contains  $X$}   \\     
			$D_e=X_e $ &\makecell[c]{the erased dataset $D_e$ contains inputs $X_e$ }   \\  
			$D_r= X_r$ &\makecell[c]{the remaining data $D_r$ contains inputs $X_r$ }   \\  
			\makecell[c]{$Z$} &the learned semantic representation \\  
	$ \text{x}, \text{z}$ & the persample from $X, Z$ \\ 
	$\mathcal{M}(f_{\theta}, g_{\theta})$& \makecell[c]{the semantic communication model $\mathcal{M}$ \\ with encoder $f_{\theta}$ and decoder  $g_{\theta}$ } \\  
	$\mathcal{M}_{u}$ & the unlearned model \\
	$f_{\theta}^{fix}$& the fixed encoder of a trained VIB model \\  
	$g_{\theta}^{fix}$& the fixed decoder of a trained VIB model \\  
	$\mathcal{L}_{JU}$ & Joint unlearning loss\\  
	$\mathcal{L}_{CC}$ & \makecell[c]{Contrastive compensation loss} \\  
	$\beta$ & the Lagrange multiplier for training a VIB model\\   
	$\alpha_1, \alpha_2$ & the weights for $\mathcal{L}_{JU}$ and $\mathcal{L}_{CC}$ \\
	\bottomrule[0.12em]
\end{tabular}}
\end{table}

\begin{figure*}[t]
	\centering
	\includegraphics[width=0.95\linewidth]{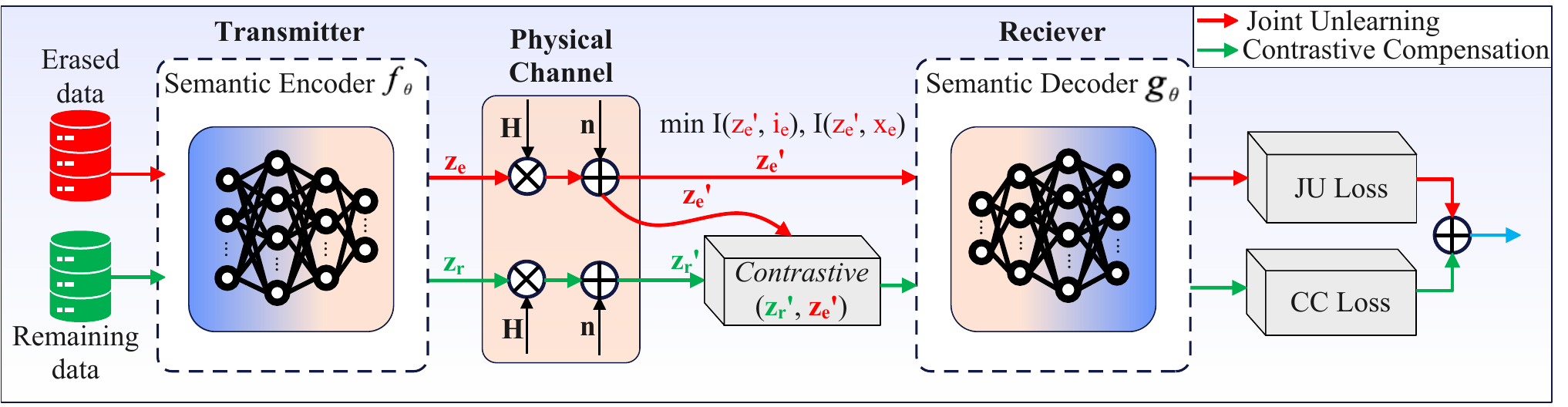}
	\vspace{-2mm}
	\caption{The overall procedure of SCU. First, we remove the influence of erased data from both semantic encoder $f_\theta$ and decoder $g_\theta$ via minimizing the mutual information between the semantic extracted representation $z_e$ and the specified sample ($i_e$); {we name the process as joint unlearning, as shown in red arrows, where $\mathbf{z}_e'$ and $\mathbf{z}_r'$ mean the representations with simulated channel noise.} Second, we retrain the unlearned models based on the remaining dataset using contrastive compensation to achieve semantic consistency, as shown in green arrows.
	}
	\label{fig_twostagesemanticunlprocess}
\end{figure*}

\subsection{Unsupervised Variational Information Bottleneck}


{The variational information bottleneck (VIB) in an unsupervised form \cite{alemi2016deep} is an important implementation to build semantic communications. It was proved that the unsupervised form VIB could be transformed into the same mathematical form as VAE \cite{kingma2014auto}. Both methods are commonly used in training semantic models \cite{hu2023robust,ma2023task,zhang2022deep}.} Since unsupervised VIB and VAE are easy to transform into each other and are easy to extend to normal encoder/decoder training, we choose to introduce our method based on the unsupervised VIB framework.

The VIB objective in an unsupervised version can be described and expanded below
\begin{equation} \label{eq:total_loss_q}
	\small
		\begin{aligned} 
			\mathcal{L}_{\mathcal{IB}}^{unsup}& =   \mathcal{L}_{enc} + \mathcal{L}_{dec} \\
			                                            & =  \beta I(Z;i) - I(Z;X) \\
		 &\simeq  \beta \text{KL}[p_{f_\theta}(Z|\text{x}_i)||  \prod_{i}^{|Z|} q_i( \text{z}_i)] +\\
		   &+\frac{1}{N} \sum_{i=1}^{N} \mathbb{E}_{ \text{z} \backsim p_{f_\theta}(Z| \text{x}_i)}[-\log \ p_{g_\theta}( \text{x}| \text{z})],
		\end{aligned}
\end{equation} 
{where $I(Z;i)$ is the mutual information (MI) between the semantic representation $Z$ and input sample $i$, which is optimized by the encoder $f_{\theta}$. $I(Z;X)$ is the mutual information between the representation and the original data $X$, which is optimized by the decoder $g_\theta$.} The corresponding encoding loss function is $\mathcal{L}_{enc}=\beta I_{f_\theta}(Z;i)$ and decoding loss function is $\mathcal{L}_{dec}=-I_{g_{\theta}}(Z;X)$. To optimize them in DL, existing work \cite{alemi2016deep} proposes a variational distribution $q(Z)$ and minimizes the corresponding upper bound of \Cref{eq:total_loss_q}. The per-sample optimization form is shown as the approximate equation in \Cref{eq:total_loss_q}. In \cite{alemi2016deep}, \citeauthor{alemi2016deep} proved that the unsupervised VIB takes the same form as VAE \cite{kingma2014auto}, except with the weight $\beta$ for encoder. 


\subsection{Building Semantic Communication with Unsupervised VIB}

We implement the semantic communication models following the mainstream semantic communication framework, JSCC \cite{bourtsoulatze2019deep}, where the transmitter owns the encoder $f_\theta$ and the receiver has the decoder $g_\theta$. They communicate through a physical channel. For easy to introduce, we assume that the system has $N_t$ transmitting antennas and $N_r$ receiving antennas. Then, the encoding symbol stream of the transmitter can be described as $\mathbf{z} = f_\theta(x), \mathbf{z} \in \mathbb{C}^{N_t \times 1}$, and $f_{\theta}$ is the encoder of VIB models that we introduce above. {Subsequently, the received signal can be described as $\mathbf{z'} = \text{H}\mathbf{z} + \mathbf{n}$, where $\text{H} \in \mathbb{C}^{N_r \times N_t}$ denotes the channel matrix and $\mathbf{n} \in \mathbb{C}^{N_r \times 1} \sim \mathcal{CN}(0,\sigma^2\text{I})$ is the additive white Gaussian noise (AWGN).} The notation $\text{I}$ denotes an identity matrix. Correspondingly, the decoding stream of the receiver can be represented as $\hat{ \text{x} } = g_\theta(\mathbf{z'})$. The procedure to train a semantic communication system is similar to the joint training process as illustrated in the upper half of \Cref{fig_twostagesemanticunlprocess}, but with an unsupervised VIB loss function \Cref{eq:total_loss_q}.

\subsection{Problem Statement} \label{ps}

{
In a semantic communication framework, assuming implemented using the VIB method, the difference between the traditional unlearning scenarios is that there are two DL models, one encoder $f_{\theta}(\cdot)$ and one decoder $g_{\theta}(\cdot)$.} The encoder is designed to extract semantic information from source inputs, while the decoder focuses on reconstructing the transmitted data from received signals. Both models undergo joint training prior to their deployment in communication. 
For convenience to understand, we assume the trained semantic communication models as a big model $\mathcal{M}(f_{\theta}, g_{\theta})$ using the unsupervised VIB method based on the whole dataset $D$. When the unlearning request about the erased dataset $D_e$ comes, the problem of semantic communication unlearning can be stated as:
{
\begin{prob_state}[Semantic Communication Unlearning] \label{eq_eu_with_C}
	Suppose there is an already-trained semantic communication system $\mathcal{M}(f_{\theta}, g_{\theta})$ based on the entire collected data $D$ using the unsupervised VIB algorithm $\mathcal{IB}$.
	Let $D_e=X_e$ be the erased dataset of an unlearned user and $D_e \subseteq D$, and we use $D_r$ to denote the remaining dataset except $D_e$. 
	The goal of the semantic communication unlearning is to design an unlearning method $\mathcal{U}$ that simultaneously unlearns both encoder $f_\theta$ and decoder  $g_\theta$ and makes the unlearned semantic model equal to the retrained model:
	\begin{equation} \small
		P(\mathcal{M}_u(f_{\theta}, g_{\theta})\!= \!\mathcal{IB}(D_r) )\!=\!P(\mathcal{M}_u(f_{\theta}, g_{\theta})\!=\!\mathcal{U}(D_e, \mathcal{IB}(D))  ).
	\end{equation}
\end{prob_state}

The key challenge of the problem lies in designing unlearning methods customized for both encoder $f_\theta$ and decoder $g_\theta$ jointly, which is much different from unlearning the normal supervised ML scenarios. During unlearning semantic models, we need to maintain the former learned semantic knowledge consistent in both codecs and eliminate the information of the erased samples from the IB-based semantic systems.}



\section{Semantic Communication Unlearning} \label{sec_SCU}

\subsection{Overview of SCU}


The key procedure of the proposed method is illustrated in Figure \ref{fig_twostagesemanticunlprocess}, encompassing two main components: joint unlearning and contrastive compensation.

\vspace{2mm}
\noindent
\textbf{Joint Unlearning.}
Joint unlearning aims to erase the contribution of specified data from the trained encoder and decoder by minimizing the mutual information between the learned semantic representation and the erased data. Joint unlearning can already implement unlearning and achieve a similar or slightly better effect than baseline methods that apply state-of-the-art unlearning methods in semantic communications. 

\vspace{2mm}
\noindent
\textbf{Contrastive Compensation.}
To further reduce the degradation of utility due to unlearning, we have developed a contrastive compensation step, which retrains the unlearned models using a contrastive loss of the erased and remaining normal semantic representations. By executing these two-step operations, SCU can effectively unlearn the semantic encoder and decoder, meanwhile preserving the semantic knowledge learned before.


\subsection{Joint Unlearning}

We first introduce how the joint unlearning in SCU implements the removal of the specified data from the trained semantic encoder and decoder. The core concept of joint unlearning involves minimizing the mutual information $I(Z;i_e)$ between the erased sample ($i_e$) and the semantic representation $Z$. This step aims to remove the erased data's influence from the encoder. Additionally, minimizing the mutual information $I(Z; X_e)$ between $Z$ and the original recovery target $X_e$ is crucial for unlearning the decoder. The ideal unlearned encoder and decoder contain no information of erased data $D_e$, i.e., $I(Z; i_{e})=0$ and $I(Z; X_e)=0$. Directly minimizing the mutual information terms $I(Z; i_e)$ and $I(Z; X_e)$ can effectively facilitate the desired unlearning process. However, this approach risks causing the representations to lose all previously acquired knowledge, not just the specific information related to the erased data. To tackle this problem, we introduce a constraint based on the Kullback-Leibler divergence (KLD) to ensure that the encoder and decoder, after unlearning, remain closely aligned with the original semantic models, thus minimizing deviation during the unlearning process. In particular, the loss of the encoder unlearning can be described as 
\begin{equation} \label{en_unl}
	\mathcal{L}_{enc}^u =   \underbrace{I_{f_{\theta}}(Z;i_e)}_{\text{MI minimization term}} + \underbrace{\text{KL}[p_{f_{\theta}}(Z|i_e) ||p_{f_{\theta}^{fix}}(Z|i_e)]}_{\text{KLD limitation term}},
\end{equation}
and the loss to unlearn the decoder can be described as 
\begin{equation} \label{de_unl}
	\mathcal{L}_{dec}^u =  \underbrace{I_{g_{\theta}}(Z;X_e)}_{\text{MI minimization term}}  + \underbrace{\text{KL}[p_{g_{\theta}}(X_e|Z)||p_{g_{\theta}^{fix}}(X_e|Z)]}_{\text{KLD limitation term}},
\end{equation}
where $f_{\theta}^{fix}$ and $g_{\theta}^{fix}$ represent the trained encoder and decoder that are fixed to compute the KLD limitation term. 
{ 
To calculate the KLD, in experiments, we keep two models: the fixed model $\mathcal{M}(f_{\theta}^{fix}, g_{\theta}^{fix})$ and the unlearned model $\mathcal{M}_u(f_{\theta}, g_{\theta})$. Both are initialized using the original trained model. During the unlearning process, we feed the same data into both fixed and unlearned models to calculate the KLD, and we only update the unlearned model.
} 
This ensures the unlearned models are not too different from the original encoder and decoder. By integrating these two loss functions, we formulate the joint unlearning loss as 
\begin{equation}\label{ju_loss}
	\begin{aligned}
		\text{JU Loss:}   \hspace{0.4cm}	\mathcal{L}_{JU} &= \mathcal{L}_{enc}^u + \mathcal{L}_{dec}^u. 
	\end{aligned}
\end{equation}
Minimizing the loss Eq. \eqref{ju_loss} based on the erased data and original trained encoder and decoder can be proved consistent with retraining the semantic models using the unsupervised VIB method based on the remaining dataset. We will present it in the theoretical analysis section.


\subsection{Contrastive Compensation}

The contrastive compensation is shown in Figure \ref{fig_twostagesemanticunlprocess} as the green arrows. 
This step is designed to achieve two purposes: preserving the semantic knowledge consistency learned previously on the remaining data; further removing the impact of the erased samples from the semantic communication models. 
In this step, we retrain the unlearned encoder and decoder with the erased dataset $D_e$ and partial samples of the remaining dataset $D_r$. The contrastive compensation contains a designed contrastive loss for the encoder and the original decoding loss of the unsupervised VIB for the decoder. 

Our idea is inspired by supervised contrastive learning \cite{khosla2020supervised}, which leverages label information to construct positive and negative contrastive samples to boost model utility. In our scenario, we directly have the contrastive negative samples: the erased data, $D_e$. We select partial samples in the remaining dataset, denoted as $A(i)$, as contrastively positive samples. For the efficiency of unlearning, we do not need to retrain the models based on the whole remaining dataset. We just randomly sample the data the same size as the erased samples. The contrastive compensation loss of the semantic communication encoder can be described as 
\begin{equation} \label{contrastive_encoder} \small
	\mathcal{L}_{enc}^{con.} = \sum_{i \in I} \mathcal{L}_{enc,i}^{con.} = \sum_{i \in I} \frac{-1}{|P(i)|} \sum_{p \in P(i)} \log \frac{\exp(\text{z}_i \boldsymbol{\cdot} \text{z}_p/\tau)}{\sum_{e \in D_e} \exp(\text{z}_i \boldsymbol{\cdot} \text{z}_e /\tau)},
\end{equation}
where $P(i) \equiv \{p\in A(i): y_p=y_i\}$ is the set of indices of all positive in the sampled auxiliary dataset $A(i)$ distinct from $i$, and $|P(i)|$ is its cardinality. The symbol $\boldsymbol{\cdot}$ represents the inner (dot) product, and $\tau$, a scalar from the set of positive real numbers $\mathcal{R}^+$, denotes the temperature parameter. We set all the samples of the erased dataset $D_e$ as the negative.

The loss of the decoder is still the same as the original trained $\mathcal{L}_{dec}$ as introduced before but is based on the encoder's contrastively outputted representation $\text{z}_i$. And the total contrastive compensation loss can be described as 
\begin{equation} \label{contras_comp}
	\text{CC Loss:}   \hspace{0.4cm}	\mathcal{L}_{CC} = \mathcal{L}_{enc}^{con.} + \mathcal{L}_{dec}.
\end{equation}
For a better performance, we can even combine both the joint unlearning loss and the contrastive compensation loss together and optimize them using a dynamic parameters as  
\begin{equation} \label{contras_comp_total}
	\mathcal{L}_{TOTAL} = \alpha_1 \cdot \mathcal{L}_{JU} + \alpha_2 \cdot \mathcal{L}_{CC}.
\end{equation}
We can optimize this total loss using grid search or multi-objective optimization \cite{sener2018multi} to achieve the best performance. For training efficiency, we set $\alpha_1$ and $\alpha_2$ fixed values as 1 in the experiments, which already achieves a significant improvement in both data erasure and model utility preservation. The pseudocode about entire SCU is presented in Algorithm~1.

\begin{algorithm}[t]
	\caption{Semantic Communication Unlearning} \label{JSCU_algorithm}
	\begin{small} 
		\BlankLine
		\KwIn{$\mathcal{M}(f_{\theta},g_{\theta})$, $D_e$, $D_r$, and epochs $E$}
		\KwOut{Unlearned semantic model $\mathcal{M}_{u}(f_{\theta_u},g_{\theta_u})$}
		Initialize the fixed temperature encoder and decoder: $f_{\theta}^{fix}, g_{\theta}^{fix} \gets f_{\theta},g_{\theta}$\\
		\For{$E$ epochs}{
			Sample a minibatch, $\{(i, x_i)\}_{i=1}^m$, from $D_e$; \\
			Sample a minibatch, $\{(j,x_j)\}_{j=1}^m$, from $D_r$; \\
			Generate $z_i \backsim p_{f_{\theta}}(\cdot|x_i), z_j \backsim p_{f_{\theta}}(\cdot|x_j)$ using the encoder;  \\
			Generate the decoder received signal with simulated channel noise: $z_i'=\text{H}z_i + n, z_j'=\text{H}z_j+n$ \\
			Calculate the joint unlearning loss $\mathcal{L}_{JU}$ using Eq. \eqref{ju_loss} based on ($i,x_i,z_i'$) and $(f^{fix}_{\theta}, g^{fix}_{\theta})$ \\
			Calculate the contrastive compensation loss $\mathcal{L}_{CC}$ based on ($i,x_i,z_i'$) and ($j,x_j,z_j'$) as Eq. \eqref{contras_comp} \\
			Update the semantic communication models, including both the encoder and the decoder, based on the gradients derived from the composite loss functions as outlined in Eq. \eqref{contras_comp_total} as 
			$ (f_{\theta}, g_{\theta}) \gets (f_{\theta}, g_{\theta}) - \eta \nabla _{(f_{\theta}, g_{\theta})} (\alpha \cdot \mathcal{L}_{JU}  + \beta \cdot \mathcal{L}_{CC} )$   \\
		}
		Return $\mathcal{M}_{u}(f_{\theta_u},g_{\theta_u}) \gets \mathcal{M}_{u}(f_{\theta},g_{\theta})$;
	\end{small}
\end{algorithm}

Upon receiving a request for unlearning, we can divide the full dataset as $D_e$ and $D_r$, and we have a trained model $\mathcal{M}(f_{\theta}, g_{\theta})$.
We establish a fixed model based on the currently trained model $\mathcal{M}(f_{\theta}, g_{\theta})$, which is used to calculate the KL divergence, as shown in Line 1 in \Cref{JSCU_algorithm}. Then we sample a minibatch of data points from $D_e$ and $D_r$, and generate corresponding representation $\text{z}'_i$ and $\text{z}'_j$ in Lines 3 to 6. We calculate the $\mathcal{L}_{JU}$ using Eq. \eqref{ju_loss} based on ($i,\text{x}_i, \text{z}_i'$) and the fixed model $(f^{fix}_{\theta}, g^{fix}_{\theta})$. And we calculate the contrastive compensation loss $\mathcal{L}_{CC}$ based on generated representations $\text{z}'_i$ and $\text{z}'_j$. Finally, we update the unlearned modal during training in Line 9.



\section{Performance Analysis} \label{theoretical} 
We have two goals of this performance analysis. First, we will prove that continuing training JU loss to unlearn a trained VIB-based semantic communication model is consistent with retraining from scratch. Second, we will analyze the time and space complexity of our method.

\subsection{Unlearning Analysis}
In the unlearning analysis for SCU, we mainly try to prove the semantic joint unlearning loss, Eq. \eqref{ju_loss}, is consistent with retraining the semantic communication system using the unsupervised VIB method based on the remaining dataset.


\vspace{2mm}
\noindent
\textbf{JU loss, Eq. \eqref{ju_loss}, is consistent with retraining a unsupervised VIB model.} 
Our method trains the original semantic communication codecs using the unsupervised VIB method through Eq. \eqref{eq:total_loss_q}.
We can combine the original unsupervised VIB loss and JU loss to prove that minimizing Eq. \eqref{ju_loss} based on the trained semantic codecs is equivalent to minimizing the loss of retraining the semantic models using unsupervised VIB on the remaining dataset $\mathcal{L}_r = \beta I(i_r;Z) - I(X_r;Z)$.
We show the proof example of unlearning the erased samples $(i_e)$ from the encoder. Since $i = i_r \cup i_e $ and $i_r \cap i_e = \emptyset$  and data are IID, along with the conditional independence of $i_r$ and $i_e$ given $Z$, we have $I(Z;i) = I(Z;i_r) + I(Z;i_e)$.
Since $ I(Z;i_e)\geq 0$ and $\text{KL}[p_{f_{\theta}}(Z|X_{i_e}) ||p_{f_{\theta}^{fix}}(Z|X_{i_e})]\geq 0$, we can expand original and encoder unelarning loss $\mathcal{L}_{enc} + \mathcal{L}_{enc}^u$ function as 
\begin{equation} \label{enc_u_proof}
	\small
	\begin{aligned}
		\mathcal{L}_{enc} + \mathcal{L}_{enc}^u&=  I(Z;i) +  I(Z;i_e)+  \text{KL}[p_{f_{\theta}}(Z| \text{x}_{i_e}) ||p_{f_{\theta}^{fix}}(Z|\text{x}_{i_e})] \\
		&\hspace{-0.2cm}= I(Z;i_r) + 2I(Z;i_e)+  \text{KL}[p_{f_{\theta}}(Z| \text{x}_{i_e}) ||p_{f_{\theta}^{fix}}(Z| \text{x}_{i_e})]  \\
		&\geq I(Z;i_r) 
	\end{aligned}
\end{equation} 
The sum of the encoder training loss $\mathcal{L}_{enc}$ and the encoder unlearning loss $\mathcal{L}_{enc}^u$ serves as an upper bound for the mutual information $I(Z;i_r)$. Minimizing this upper bound during original training and unlearning is simultaneously minimizing $I(Z;i_r)$ itself. 

In a similar way, we can prove that minimizing the joint unlearning loss of the decoder to unlearn the trained decoder using \Cref{de_unl} is equivalent to minimizing the upper bound of $-I(X_r;Z)$ in retraining using the unsupervised VIB.
Hence, minimizing the JU loss function for a trained semantic model is akin to minimizing the loss function when retraining an unsupervised VIB model using only the remaining dataset.

\subsection{Complexity Analysis}


In this section, we present a thorough examination of the time and space complexity associated with the SCU method. The total number of parameters within the encoder and decoder of the semantic communication model is denoted by $p$. Furthermore, the aggregate number of samples is represented by $s$, while the dimensionality of the feature vector is denoted by $k$.

\vspace{2mm}
\noindent
\textbf{Time Complexity of SCU.} 
Let $f(p)$ denote the time complexity of forward propagation. Referring to \cite{griewank2008evaluating}, the time complexity for a single backpropagation step is at most $5f(p)$. Therefore, the total complexity for computing the derivative for each training sample is $6f(p)$. Therefore, the total time complexity of a single training round of all batches of the erased dataset (denoted by $B_u$) is $6f(p) B_u$. Assume that the unlearning process consists of $T_u$ iterations, with each iteration taking $t_u$ time to complete. The total running time of variational Bayesian unlearning (VBU) \cite{nguyen2020variational}, which is conducted only based on the erased dataset, is $6f(p) B_u T_u t_u$.
Our SCU method has a contrastive compensation training based on the selected subset from the remaining dataset. In this paper, we set the selected dataset to have the same size as the erased dataset, $B_u$. Assuming the time complexity of one contrastive forward propagation is $f_c{p}$, the total running time of SCU is $(6f(p) + 6f_c(p)) B_u  T_u  t_u$.

\vspace{2mm}
\noindent
\textbf{Time Complexity of HBU Methods.} 
For the Hessian matrix-based unlearning (HBU) methods \cite{guo2019certified,sekhari2021remember}, they need to calculate the Hessian matrix, which is based on the remaining dataset and is denoted as $B_r$. We assume the time complexity of computing the Hessian matrix of the model is $h(p)$. Therefore, their running time is $(6f(p)  B_u + h(p)  B_r)  T_u  t_u$.

\vspace{2mm}
\noindent
\textbf{Space Complexity of SCU.} 
In the proposed SCU, we jointly unlearn the encoder and decoder by minimizing the mutual information, which is similar to the original semantic model training process. Although we have to retrain the unlearned model contrastively, it is conducted based on the subset with the same size as the erased dataset.
Therefore, the space complexity of our method is linear complexity, i.e., $\mathcal{O}(2k  |D_e|+p)$, which is much lower than hessian-matrix-based methods that need quadratic complexity to calculate the Hessian matrix.

\section{Performance Evaluation} \label{ex}

{We first introduce the experimental setup. Then, we conduct experiments to evaluate the impact of the Erased Data Ratio during semantic communication models' unlearning in subsection \ref{exp_edr} and assess the influence of the channels and Signal-to-Noise Ratio (${\it SNR}$) on semantic communication model performance in subsection \ref{exp_channels}. Finally, an ablation study is conducted to demonstrate the impact of the CC loss within the proposed SCU method.
}


\begin{table*}[t]
	\scriptsize
	\caption{Overview Evaluations on MNIST, CIFAR10 and CIFAR100. }
		\vspace{-2mm}
	\label{tab_total}
	\resizebox{\linewidth}{!}{
		\begin{tabular}{ccccccccccccc}
			\toprule[1pt]
			\multirow{2}{*} {Evaluation Metrics} & \multicolumn{4}{c} {MNIST, ${\it EDR}=6\%$, ${\it SNR}=5$, AWGN}& \multicolumn{4}{c} {CIFAR10, ${\it EDR}=6\%$, ${\it SNR}=20$, AWGN} & \multicolumn{4}{c} {CIFAR100, ${\it EDR}=6\%$, ${\it SNR}=20$, AWGN}  \\
			\cmidrule(r){2-5}   \cmidrule(r){6-9} \cmidrule(r){10-13}
			& Origin  		& HBU   &VBU		 	& SCU		& Origin 		& HBU 		&VBU  		& SCU & Origin 		& HBU 		&VBU  		& SCU \\
			\midrule 
			{Bac. Acc. on erased} 	& 97.72\%  & \textbf{0.06\%}   &2.11\%      & 7.28\%      & 98.53\%	  & \textbf{2.40\%}    &7.27\%	&4.17\% & 94.64\% & \textbf{0.08\%}	&0.80\% 		& 7.12\%\\
			Acc. on clean  	 & 97.17\%   &0.01\% &68.19\%    & \textbf{97.56\%}    & 67.23\%	 & 42.90\%  &47.13\% &\textbf{55.00\%}& 43.84\% & 6.68\%		&15.04\%  		& \textbf{27.84\%} \\
			MSE on erased  & 2.92 		 & 83.33   & 77.50       & \textbf{6.13 }   		  & 39.42	      & 56.76      &61.27     &\textbf{45.18}& 26.87		& 467.58 &55.61  		& \textbf{35.41} \\
			MSE on clean    & 2.79 		  & 74.90    &71.94        & \textbf{2.80 }  		& 39.24	       & 54.21       &58.08     &\textbf{43.31}& 27.09 & 443.23 		&51.29 & \textbf{33.45} \\
			Running time (s)  & 275       &13.80 &\textbf{0.31}  & 2.1         & 2120       & 106.07     & \textbf{1.60}         &2.23& 2120 		& 106.07 		&\textbf{1.88 } 		& 3.08 \\
			\bottomrule[1pt]
	\end{tabular}}
	\vspace{-2mm}
	\begin{tabbing}
		{Bac. Acc. on erased: Backdoor accuracy on erased dataset; Acc. on clean: Model accuracy on test clean dataset.}
	\end{tabbing}
\vspace{-4mm}
\end{table*}

\subsection{Experimental Setup} \label{exp_setup}
\noindent
\textbf{Datasets.} 
We have conducted extensive experiments on three representative datasets, MNIST, CIFAR10 and CIFAR100. Although all the three datasets are usually used to train classification models, we here only utilize the inputs of the three datasets for unsupervised semantic models training. 

\vspace{2mm}
\noindent
\textbf{Compared Methods.} 
We evaluate and compare the performance of SCU and two baselines. These two baselines of semantic communication unlearning are implemented using two state-of-the-art machine unlearning methods: Hessian-matrix-based unlearning (HBU) \cite{sekhari2021remember,guo2019certified} and variational Bayesian unlearning (VBU) \cite{fu2022knowledge, nguyen2020variational}. 

\vspace{2mm}
\noindent
\textbf{Evaluation Method for SCU.} 
To effectively evaluate SCU and the compared methods, we conduct experiments to unlearn the backdoored semantic communication encoder and decoder, i.e., the models are trained based on backdoored samples, and the purpose of unlearning is to erase the backdoor from the trained models. Our evaluation method is inspired by a commonly used traditional unlearning evaluation method~\cite{hu2022membership}. {We choose the backdoor-based evaluation method rather than membership inference attack (MIA) methods because backdoor accuracy is more oblivious than MIA to evaluate if the sample is unlearned from the model \cite{chen2021machine}.}

Specifically, we first design a patch block on the right bottom of images that will be erased. The original semantic encoder and decoder are trained using these backdoored samples, leading to the learning of backdoor information. Subsequently, a downstream classification model is trained on the output of the decoder. This model learns the backdoor information embedded in the recovered samples, which is present due to the decoder having learned this backdoor information. We verify whether the decoded backdoored samples can still attack the downstream classifying model after unlearning.

\vspace{2mm}
\noindent
\textbf{Metrics.} 
SCU and the other two baselines are assessed in terms of both unlearning effectiveness and efficiency. Unlearning effectiveness includes the evaluation of semantic models' performance and downstream classifying models' performance. We summarize the following four metrics:
\begin{itemize}
	\item \textbf{MSE.} We use mean square error (MSE) to evaluate the decoding effectiveness. {Good unlearning methods for an unsupervised model should (1) keep a small decoding MSE on the clean dataset (preserving model utility) and (2) increase the MSE on the erased dataset (achieving the unlearning effect) after unlearning.} 
	\item \textbf{Model Accuracy.} 	Model accuracy is used to evaluate the downstream models' performance, which predicts based on the decoding samples of semantic models, reflecting the semantic models' utility to some extent. 
	\item  \textbf{Backdoor Accuracy.} Backdoor accuracy is calculated by the prediction of downstream models based on the decoded erased samples, which evaluates if the backdoor information is still contained in the decoder.
	\item  \textbf{Running Time.} This metric evaluates efficiency by tracking the duration of processing each training batch and then multiplying this time by the total number of training epochs.
\end{itemize}



\vspace{2mm}
\noindent
\textbf{Implemental Details.} 
Since our semantic communication models are trained using the unsupervised VIB structure, we train the encoder and decoder with two linear models on MNIST, each with three hidden layers, with a learning rate $\eta=0.0001$. And the downstream classifying model on MNIST is also a linear model. On CIFAR10 and CIFAR100, we train two Resnet18 models as the encoder and decoder with $\eta=0.0005$. The downstream classifying model on CIFAR is a Resnet18, which is trained using the recovered $\hat{x}$. During the training phase, we introduce simulated noise originating from AWGN, Rayleigh, and Rician channels. The minibatch size is 16 on all datasets. All models are developed using PyTorch, and the experiments are conducted on a computing cluster equipped with four NVIDIA 1080ti GPUs. 


\begin{figure*}[t]
	\centering
	\vspace{-4mm}
	\hspace{-4mm}
	\subfloat[\footnotesize Accuracy on clean data ]{ \label{fig_mnistaccercurve} \rotatebox{90}{ \hspace{5mm}	\scriptsize{ On MNIST} }
		\includegraphics[scale=0.17]{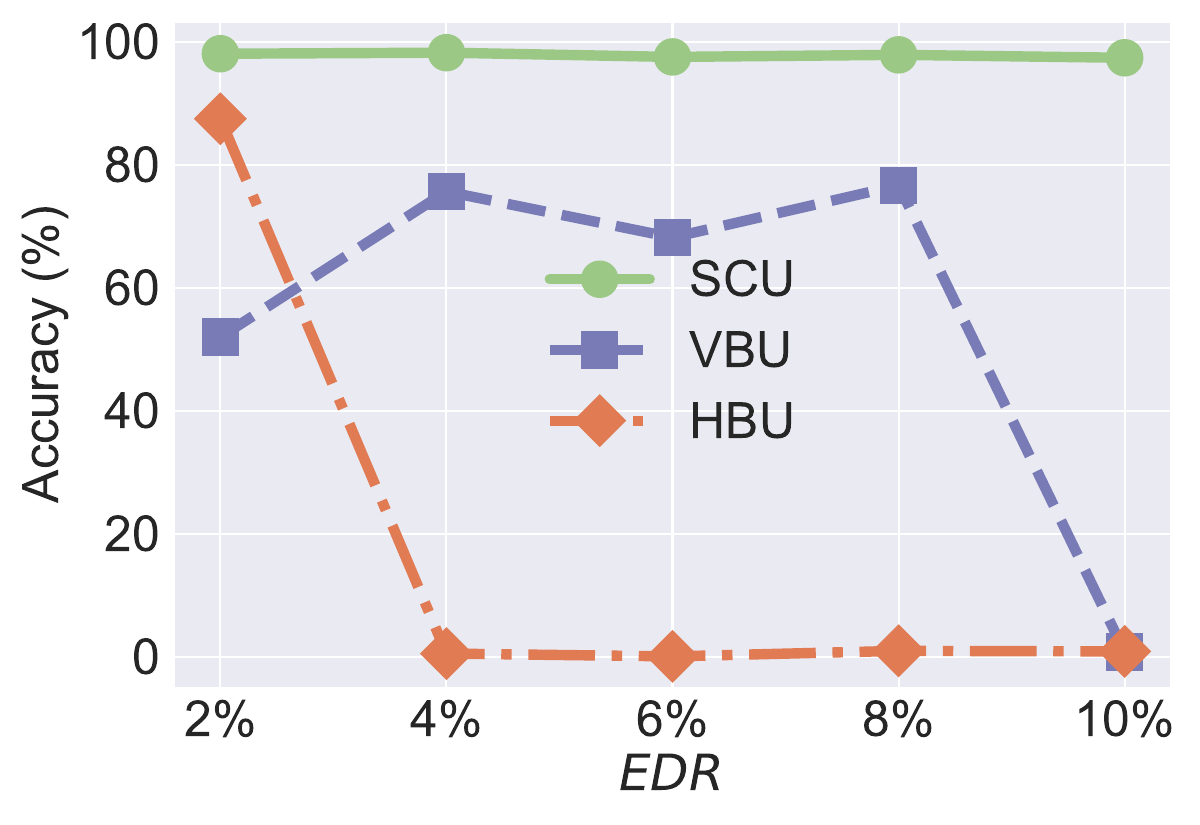}
	}\hspace{-4mm}
	\subfloat[\footnotesize Bac. acc. on erased data]{ \label{fig_mnistbackaccercurve}
		\includegraphics[scale=0.17]{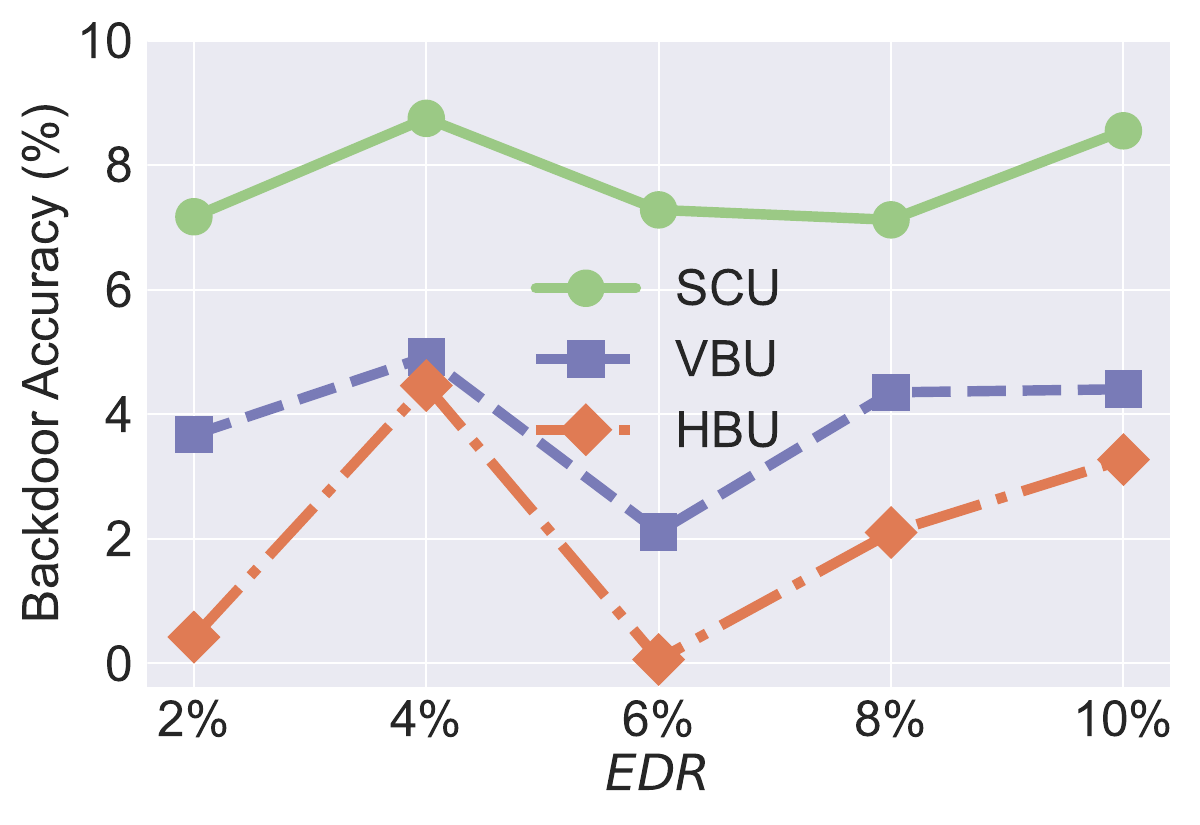}
	}\hspace{-4mm}
	\subfloat[\footnotesize MSE on clean data]{\label{fig_mnistmsecleanercurve}
		\includegraphics[scale=0.17]{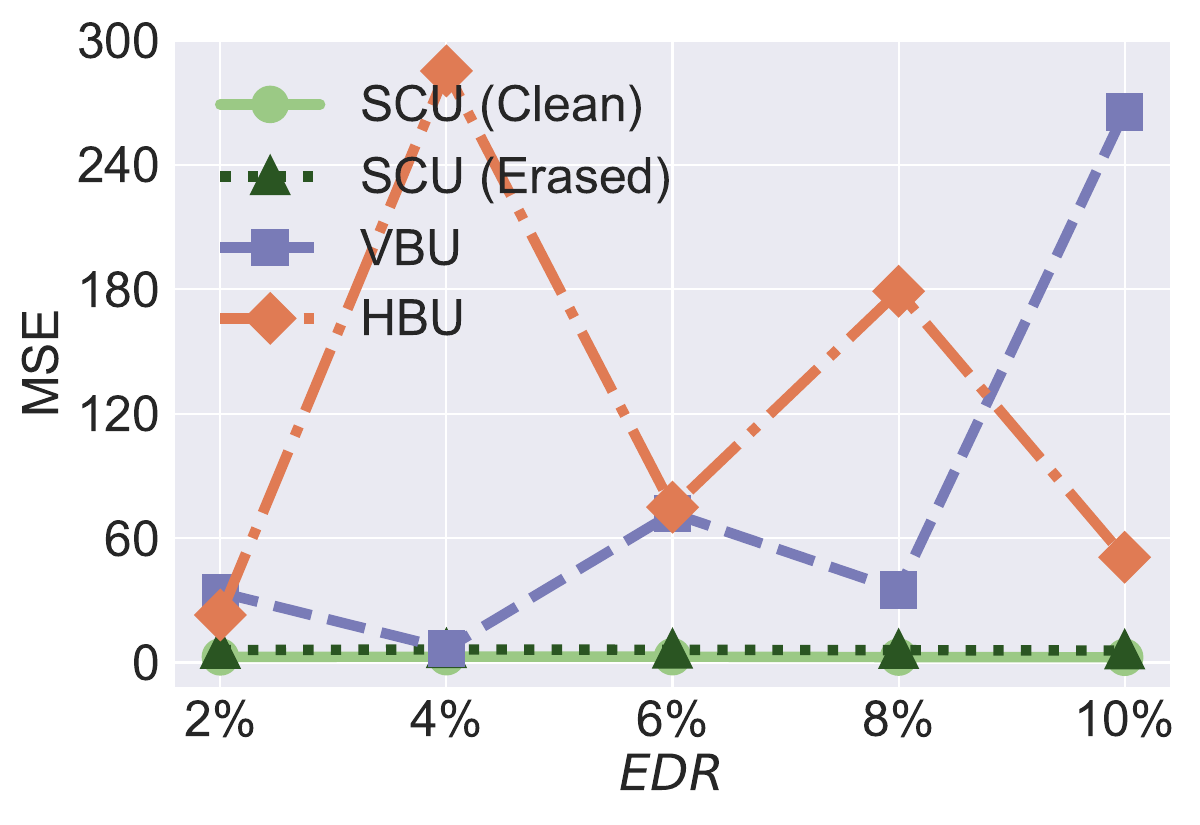}
	}\hspace{-4mm}
	\subfloat[\footnotesize MSE on erased data]{	\label{fig_mnistmseerasedercurve}
		\includegraphics[scale=0.17]{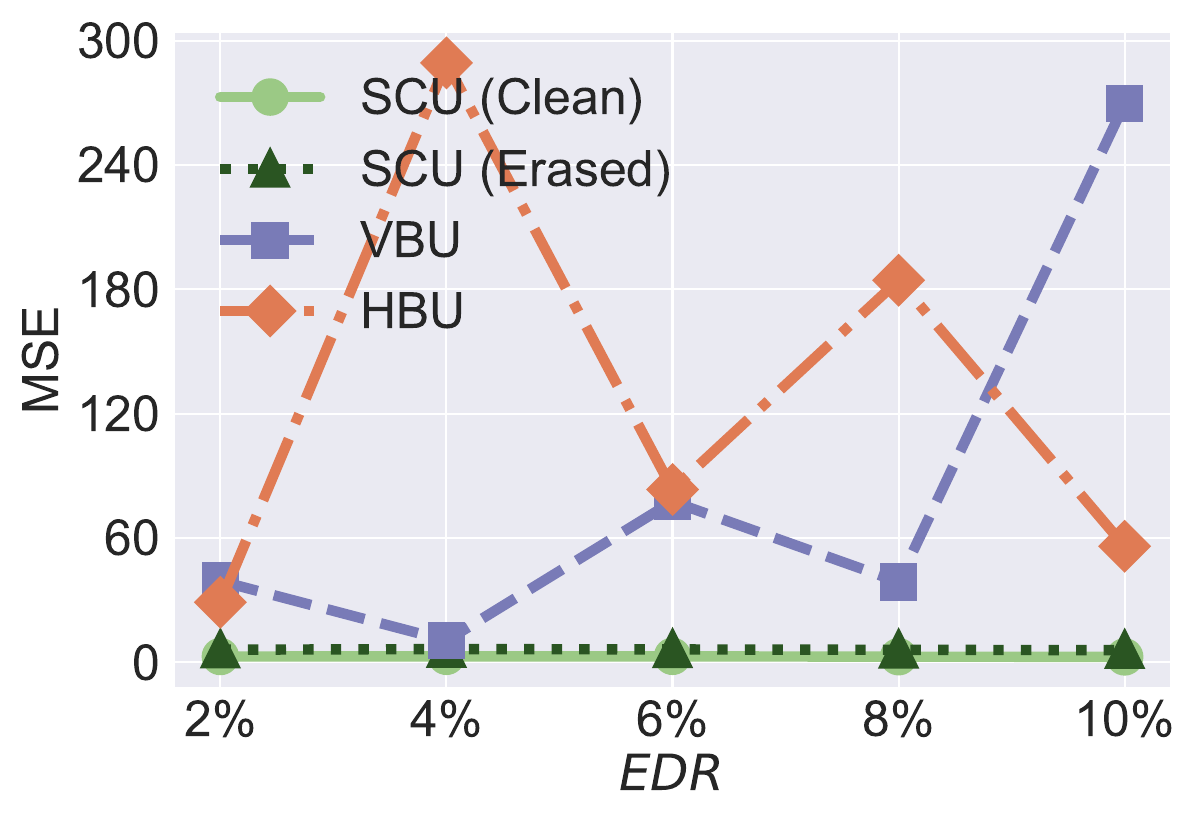}
	}\hspace{-4mm}
	\subfloat[\footnotesize Running time]{  	\label{fig_mnistrterbar}
		\includegraphics[scale=0.17]{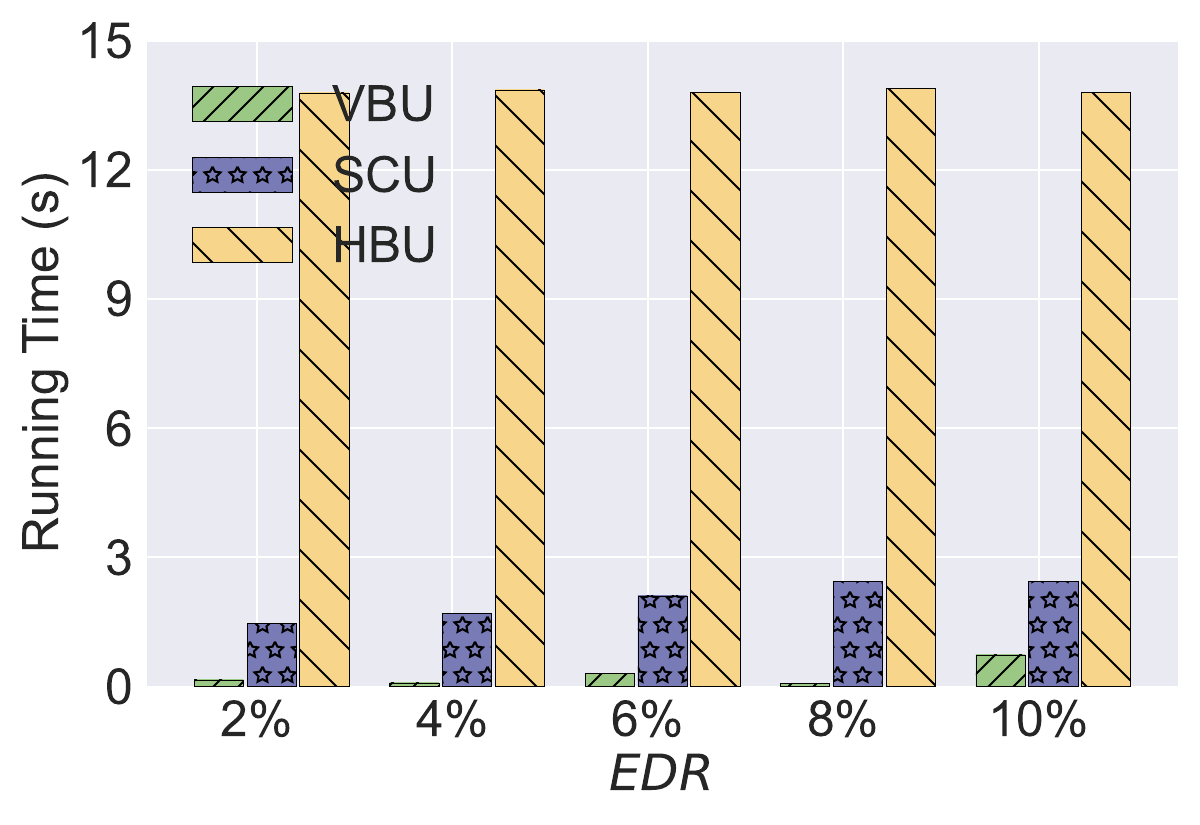}
	}\vspace{-2mm}
	\\
	\hspace{-4mm}
	\subfloat[\footnotesize Accuracy on clean data]{ 	\label{fig_cifaraccercurve} \rotatebox{90}{ \hspace{3mm}	\scriptsize{ On CIFAR10} }
		\includegraphics[scale=0.17]{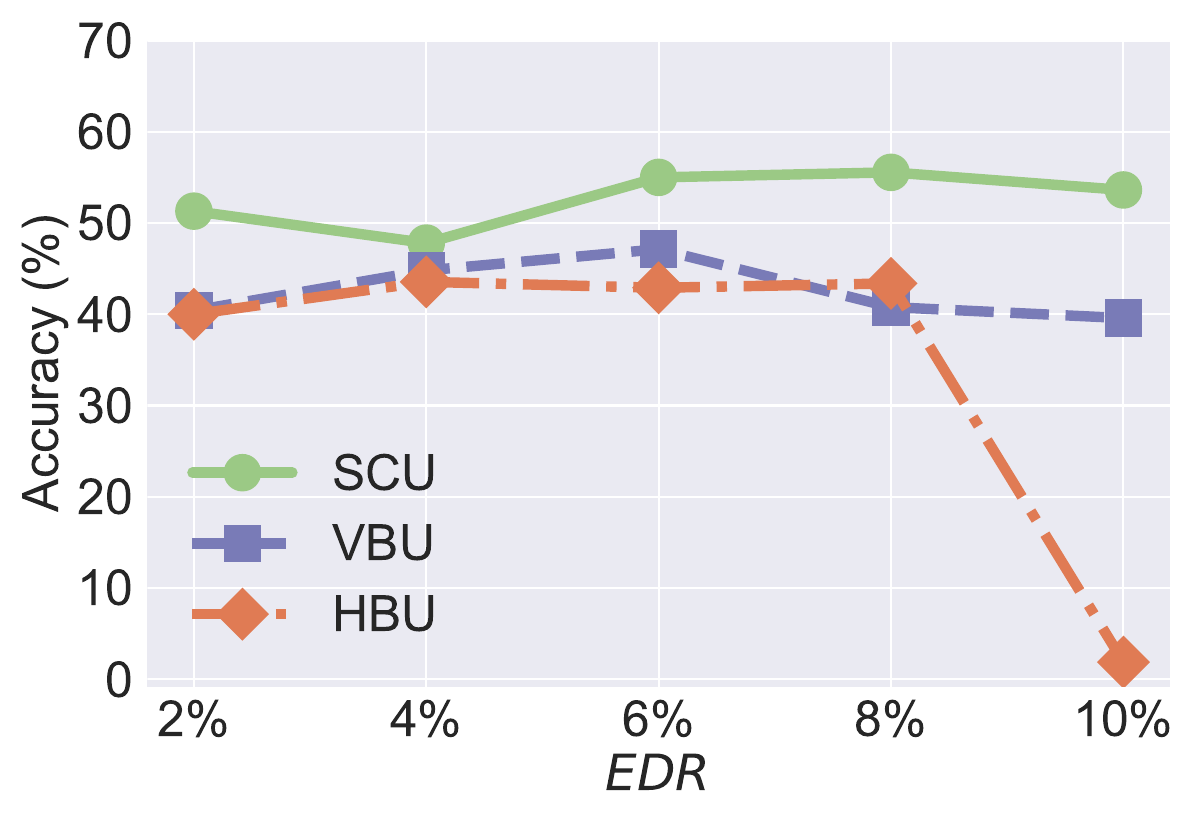}
	}\hspace{-4mm}
	\subfloat[\footnotesize Bac. acc. on erased data]{ \label{fig_cifarbackaccercurve}
		\includegraphics[scale=0.17]{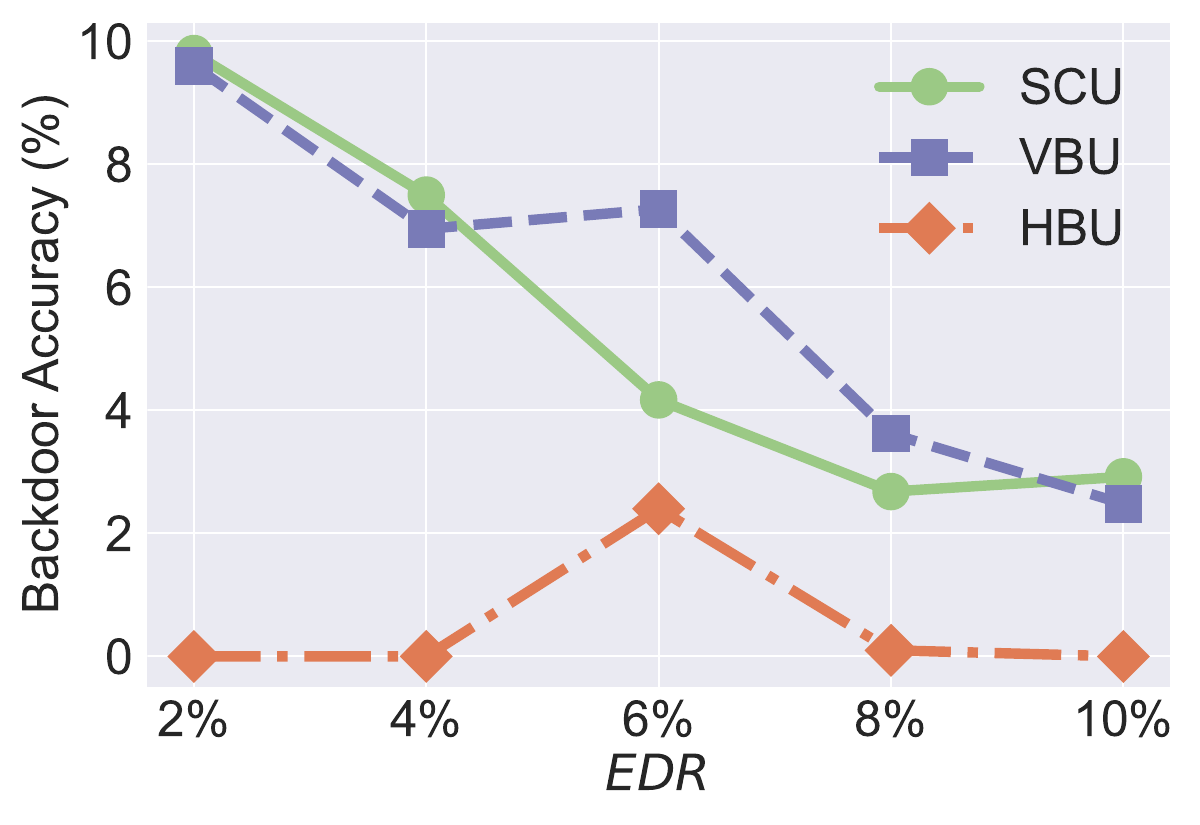}
	}\hspace{-4mm}
	\subfloat[\footnotesize MSE on clean data]{ \label{fig_cifarmsecleanercurve}
		\includegraphics[scale=0.17]{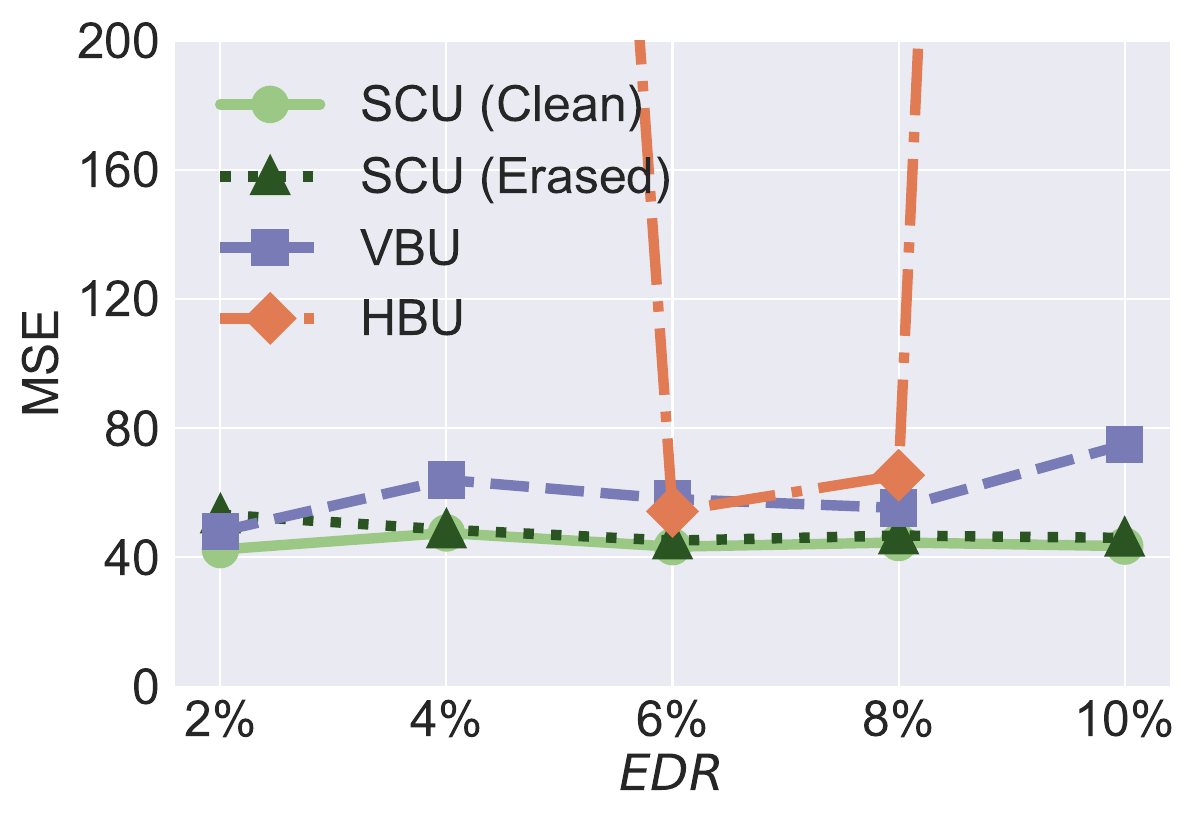}
	}\hspace{-4mm}
	\subfloat[\footnotesize MSE on erased data]{\label{fig_cifarmseerasedercurve}
		\includegraphics[scale=0.17]{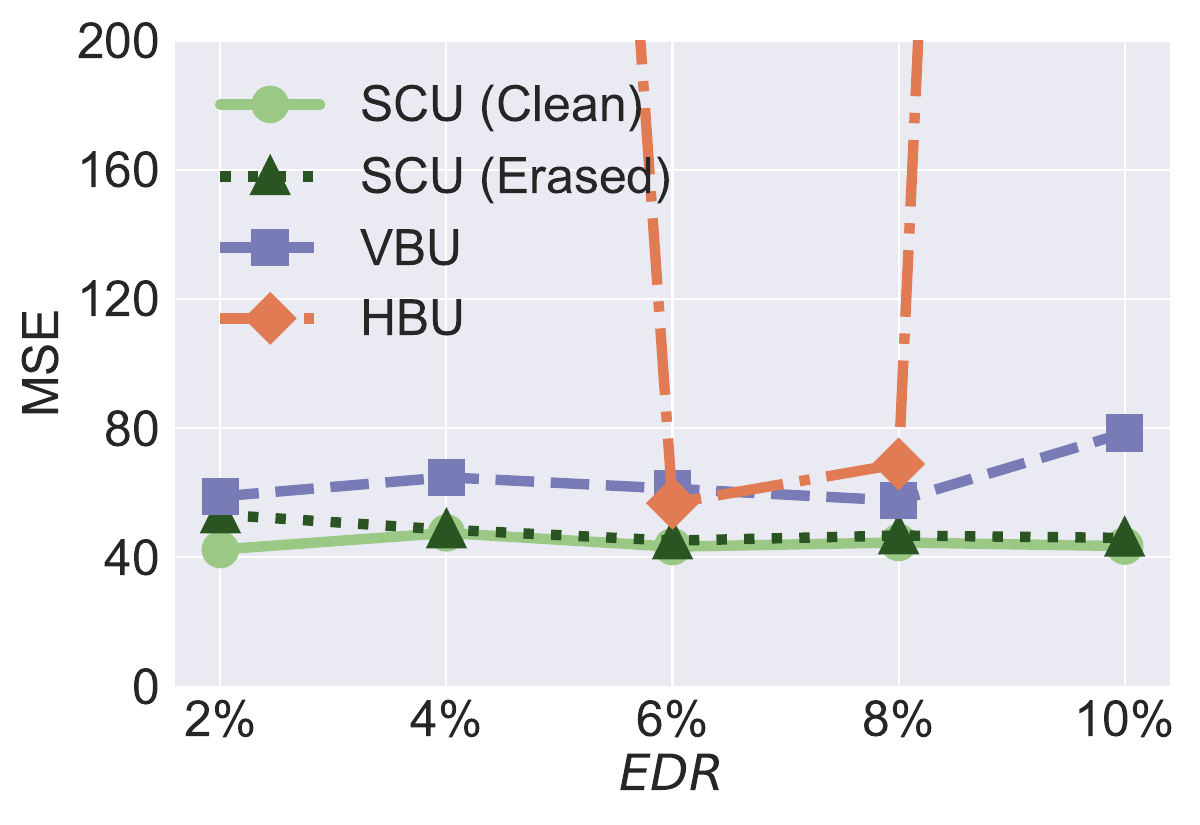}
	}\hspace{-4mm}
	\subfloat[\footnotesize Running time]{  \label{fig_cifarrterbar}
		\includegraphics[scale=0.17]{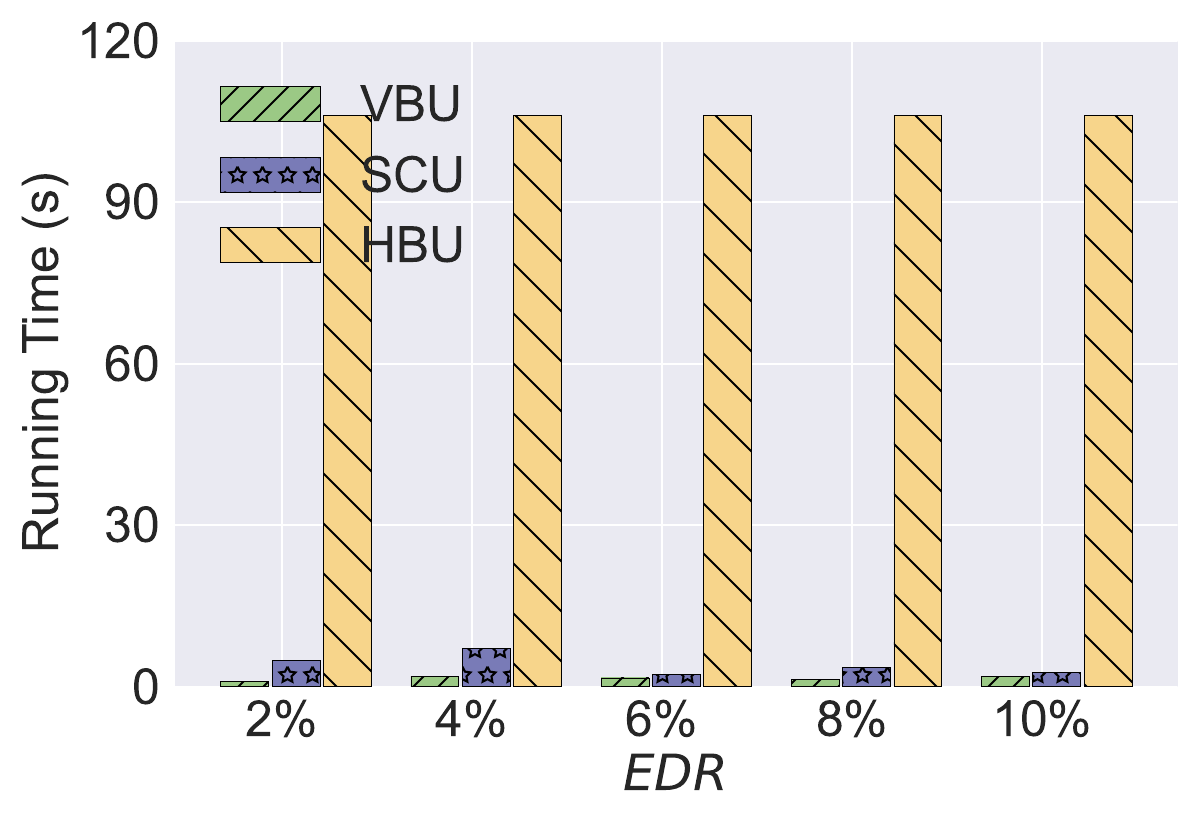}
	}\vspace{-2mm}
	\\
	\hspace{-4mm}
	\subfloat[\footnotesize Acc. on clean data]{ 	\label{fig_cifar100accercurve} \rotatebox{90}{ \hspace{3mm}	\scriptsize{ On CIFAR100} }
		\includegraphics[scale=0.17]{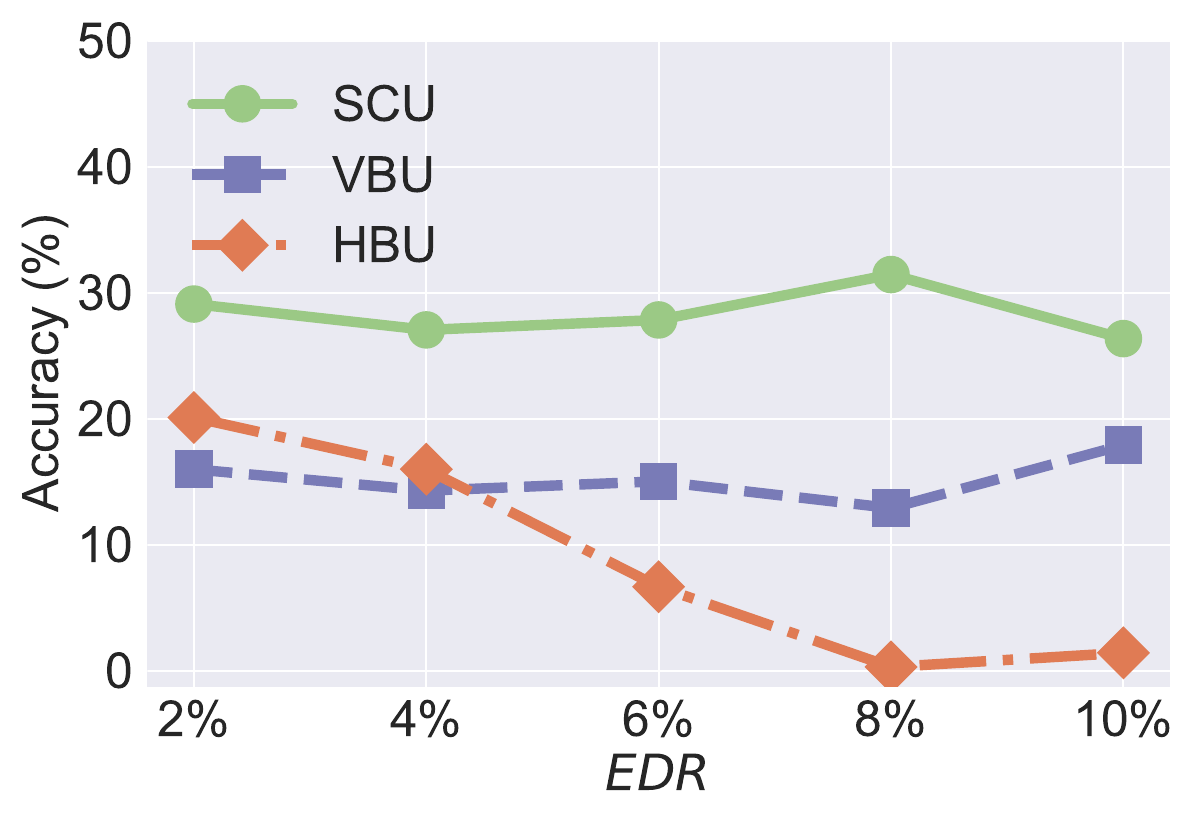}
	}\hspace{-4mm}
	\subfloat[\footnotesize Bac. acc. on erased data]{ \label{fig_cifar100backaccercurve}
		\includegraphics[scale=0.17]{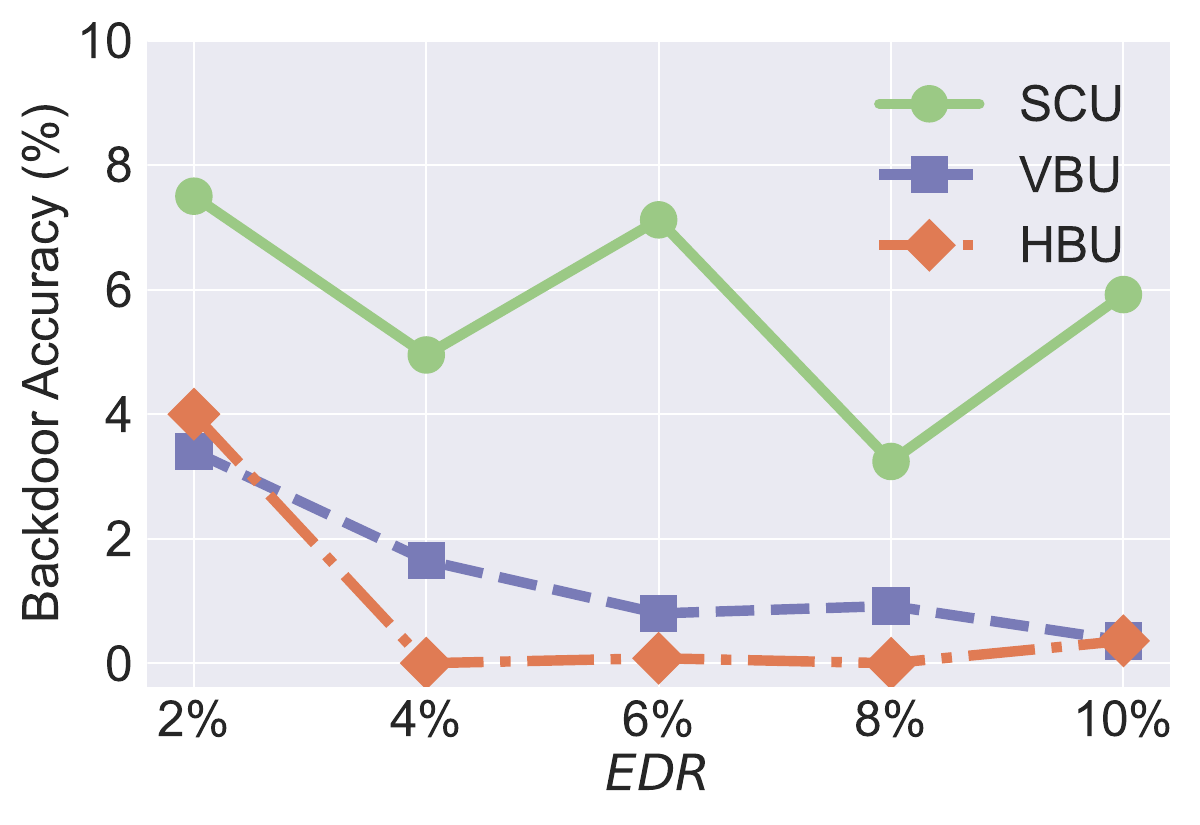}
	}\hspace{-4mm}
	\subfloat[\footnotesize MSE on clean data]{ \label{fig_cifar100msecleanercurve}
		\includegraphics[scale=0.17]{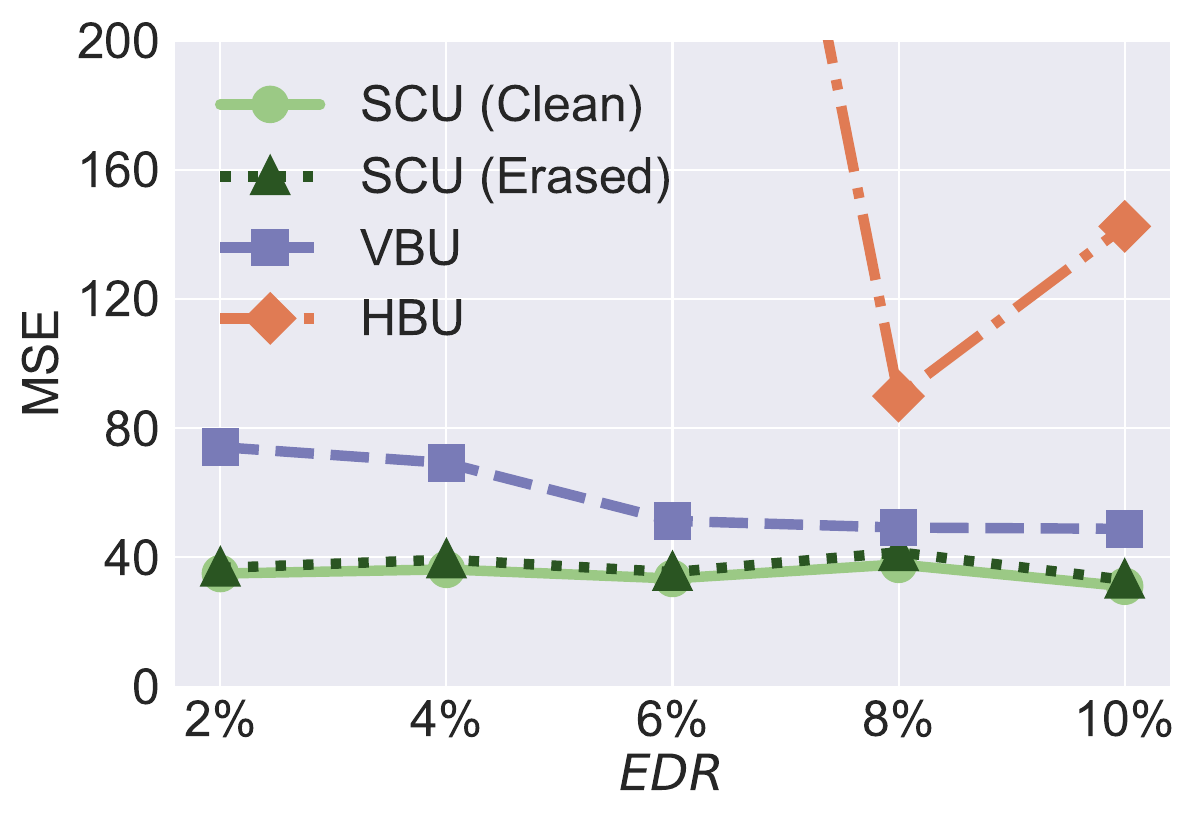}
	}\hspace{-4mm}
	\subfloat[\footnotesize MSE on erased data]{\label{fig_cifar100mseerasedercurve}
		\includegraphics[scale=0.17]{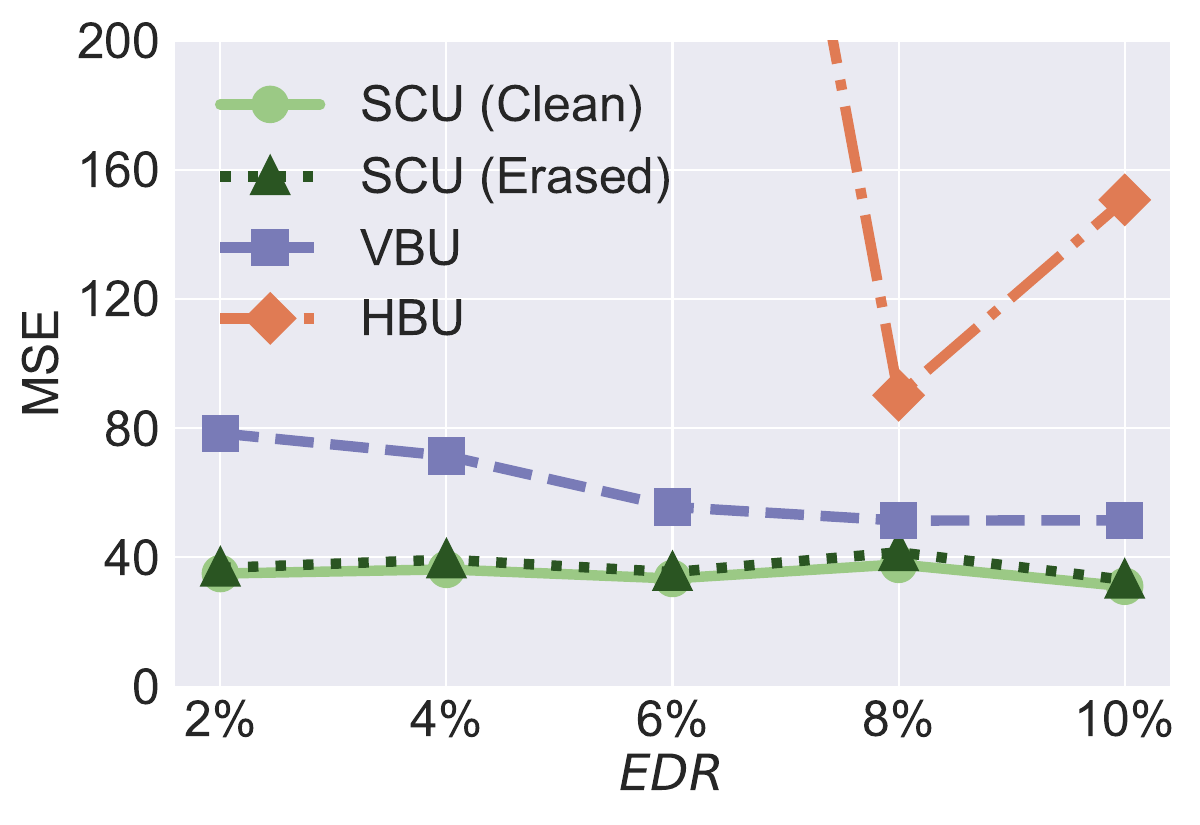}
	}\hspace{-4mm}
	\subfloat[\footnotesize Running time on]{  \label{fig_cifar100rterbar}
		\includegraphics[scale=0.17]{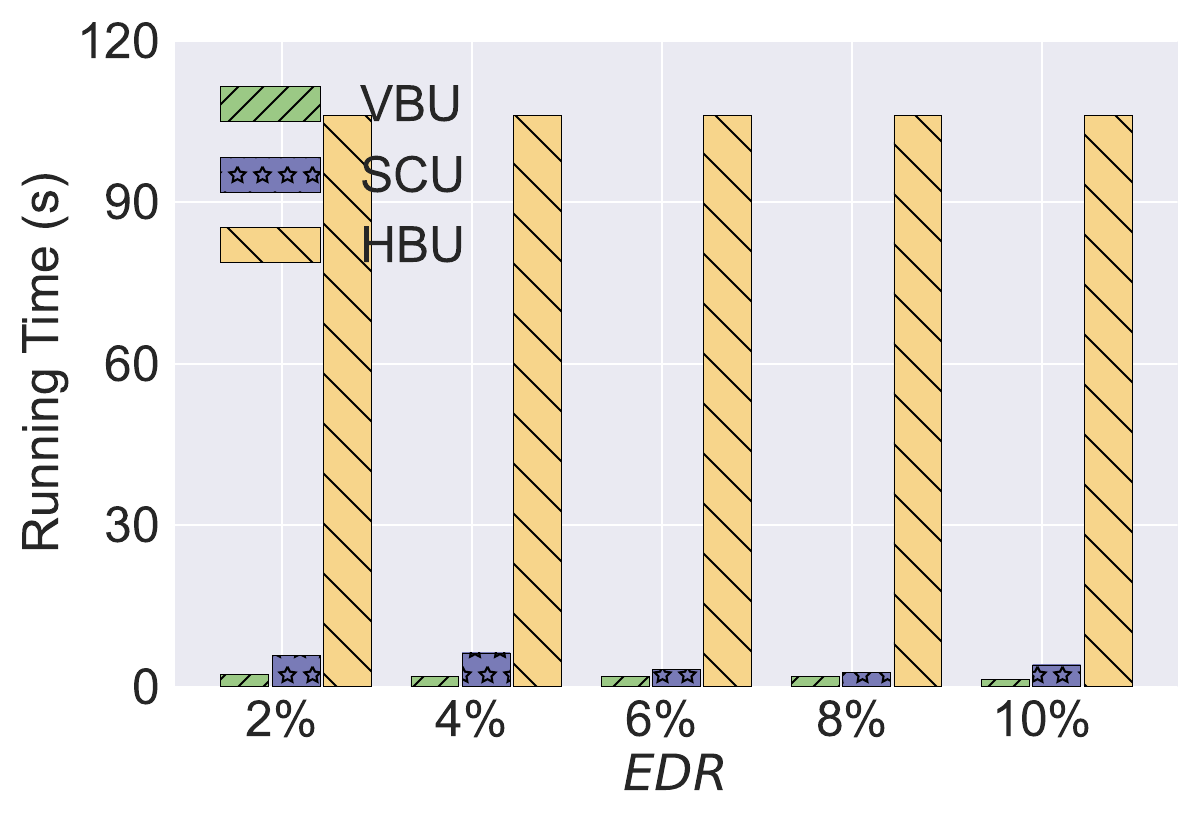}
	}
	\caption{Evaluations of the impact of different ${\it EDR}$. Model accuracy and backdoor accuracy of downstream models in SCU and VBU decrease as the ${\it EDR}$ increases. HBU easily causes catastrophic unlearning, reflected in huge model accuracy degradation and the highest decoding MSE. Especially in CIFAR10 and CIFAR100, the decoding MSE of HBU is higher than the vertical axis.} 
	\label{evaluation_of_edr} 
\end{figure*}

\begin{figure*}[!b] 
	\vspace{-0.5cm}
	\centering
	\subfloat[\footnotesize Accuracy and backdoor acc, AWGN]{ \label{fig_mnistaccsnrawgncurve}
		\includegraphics[scale=0.28]{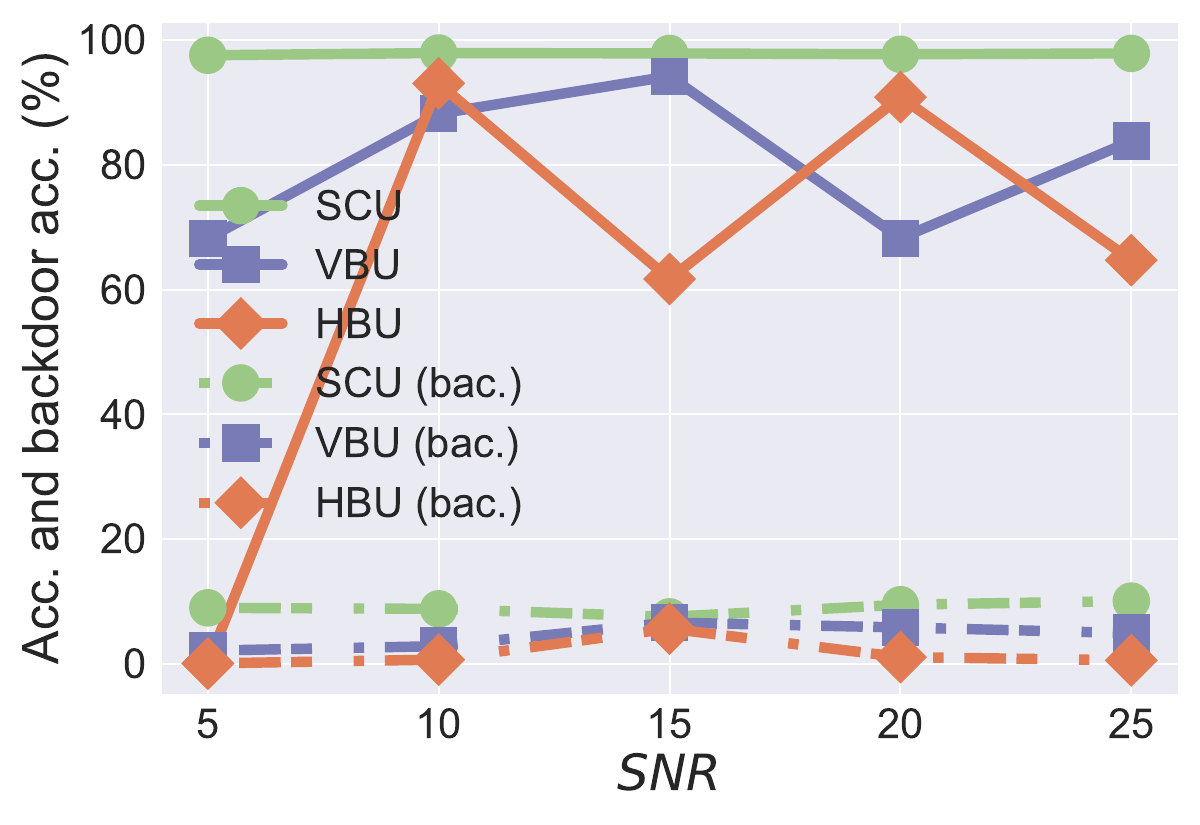}
	}
	\subfloat[\footnotesize Accuracy and backdoor acc, Rayleigh]{ \label{fig_mnistaccsnrrayleighcurve}
		\includegraphics[scale=0.28]{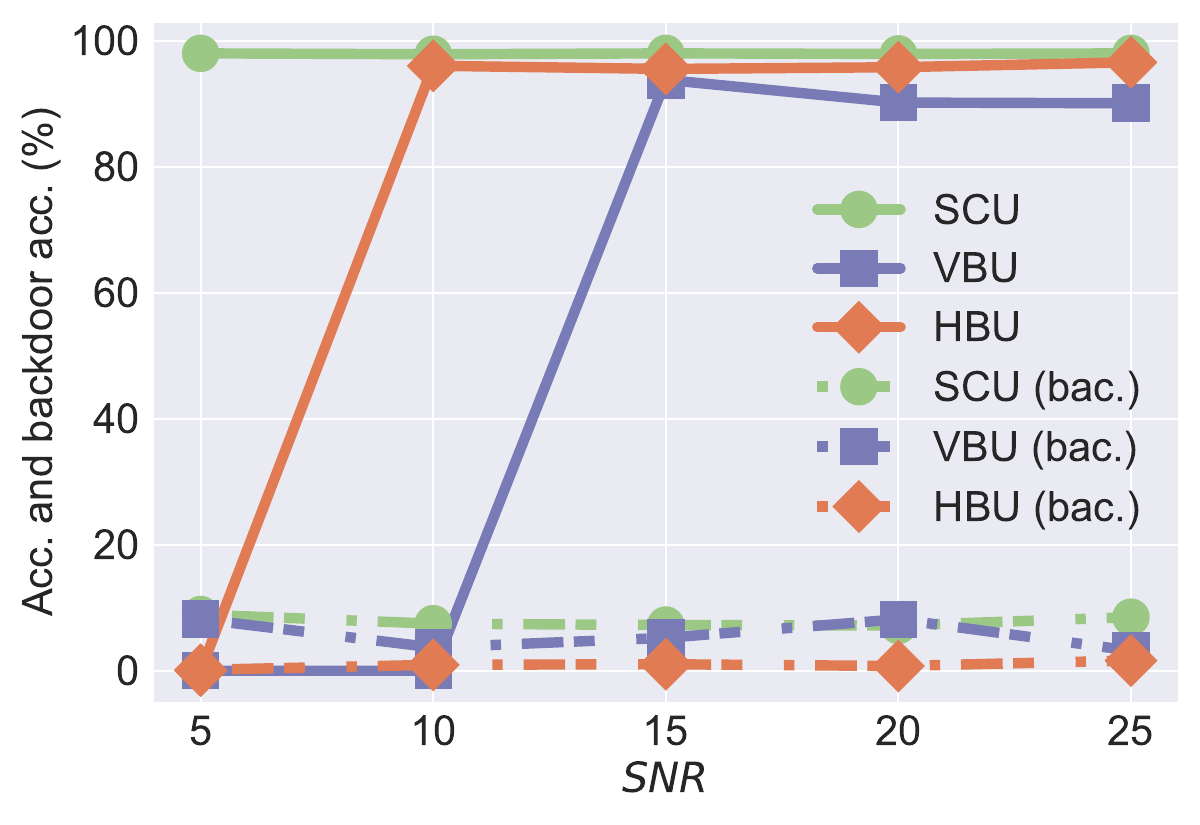}
	}
	\subfloat[\footnotesize Accuracy and backdoor acc, Rician]{ \label{fig_mnistaccsnrriciancurve}
		\includegraphics[scale=0.28]{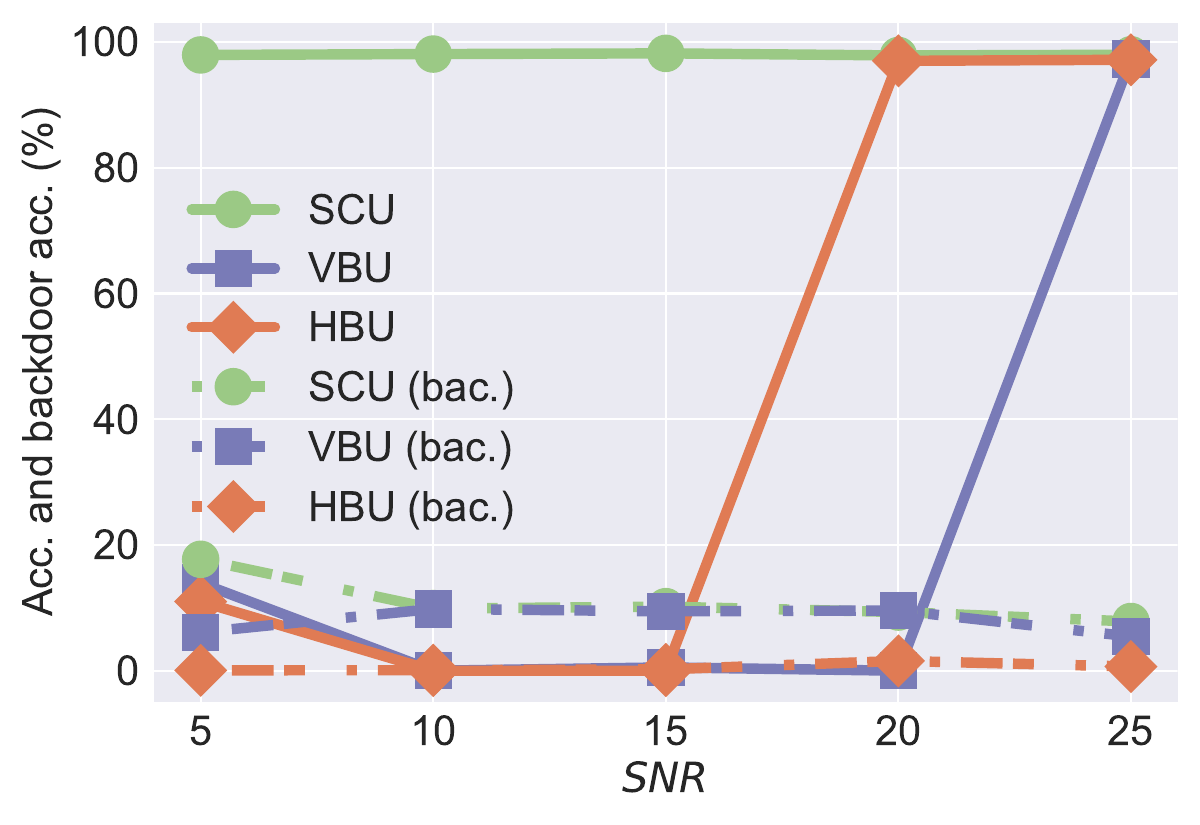}
	}
	\\ 
	\subfloat[\footnotesize MSE on clean and erased samples, AWGN]{ 	\label{fig_mnistmsecleansnrawgncurve}
		\includegraphics[scale=0.28]{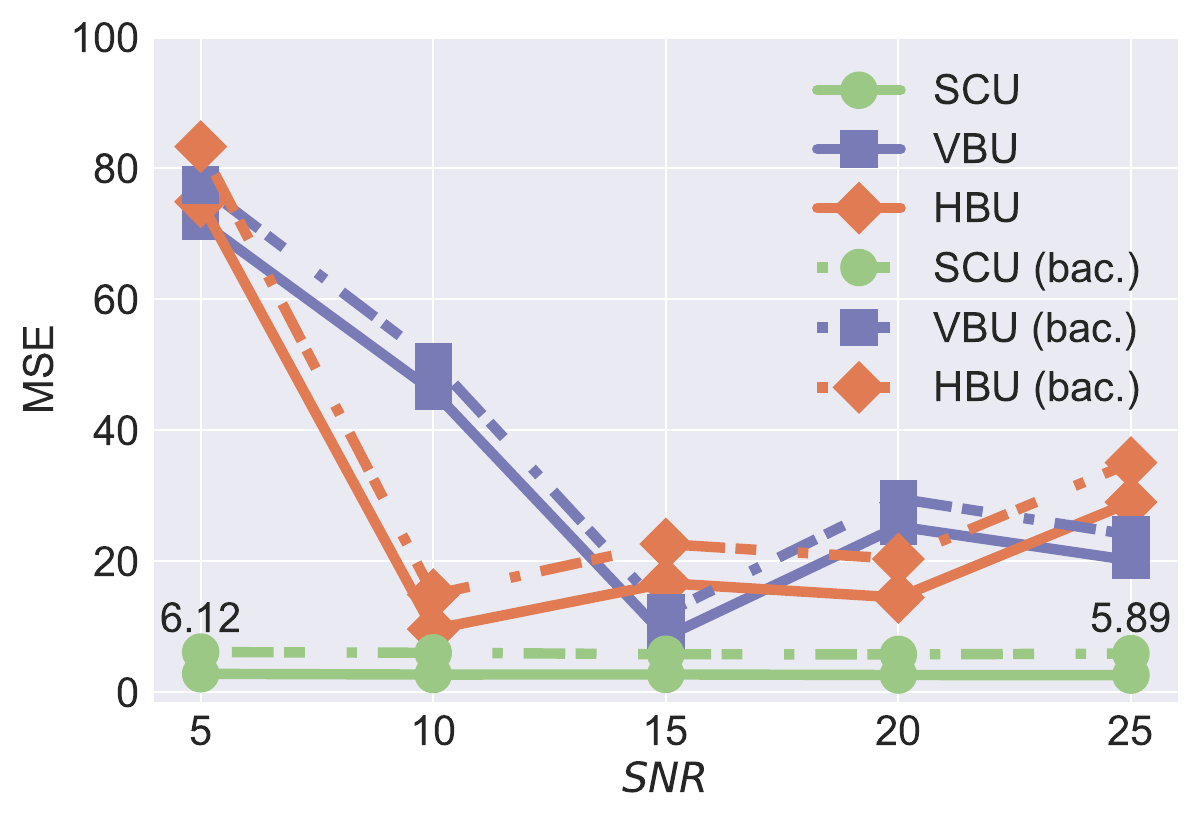}
	}
	\subfloat[\footnotesize MSE on clean and erased samples, Rayleigh]{	\label{fig_mnistmsecleansnrrayleighcurve}
		\includegraphics[scale=0.28]{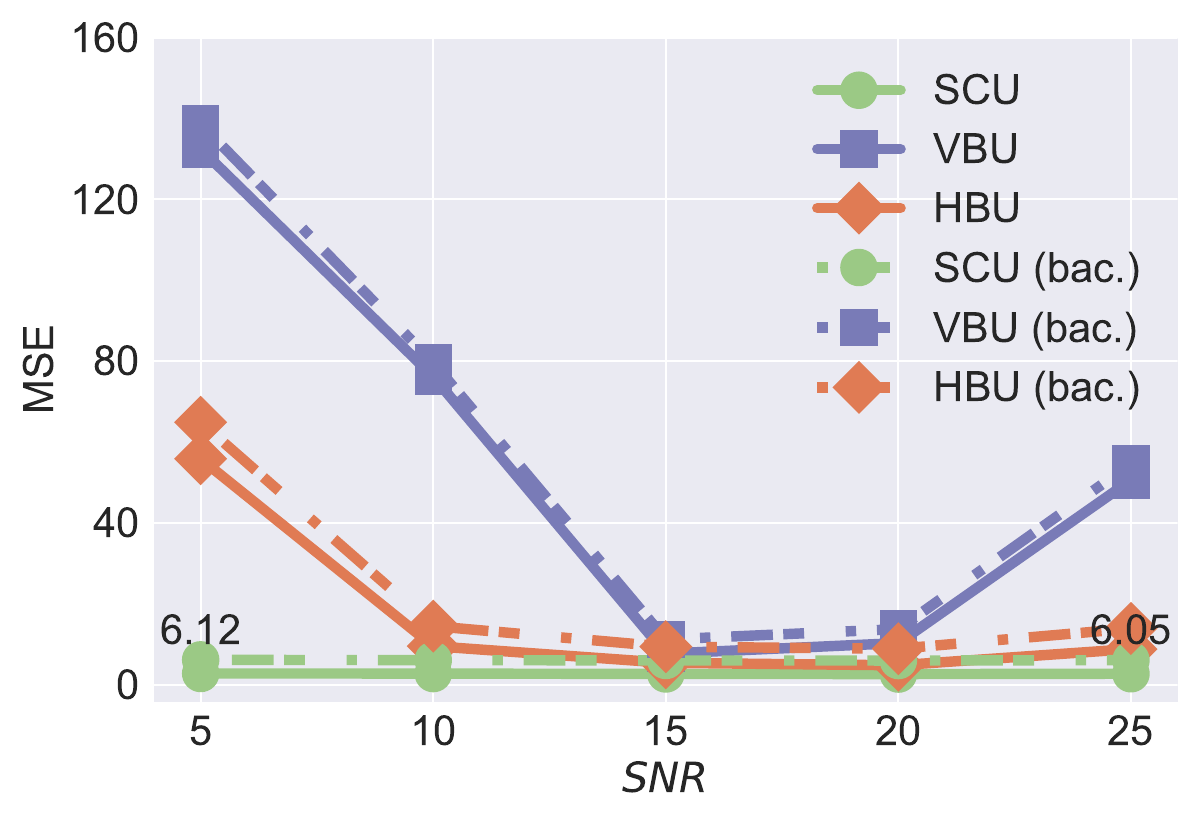}
	}
	\subfloat[\footnotesize MSE on clean and erased samples, Rician]{ 	\label{fig_mnistmsecleansnrriciancurve}
		\includegraphics[scale=0.28]{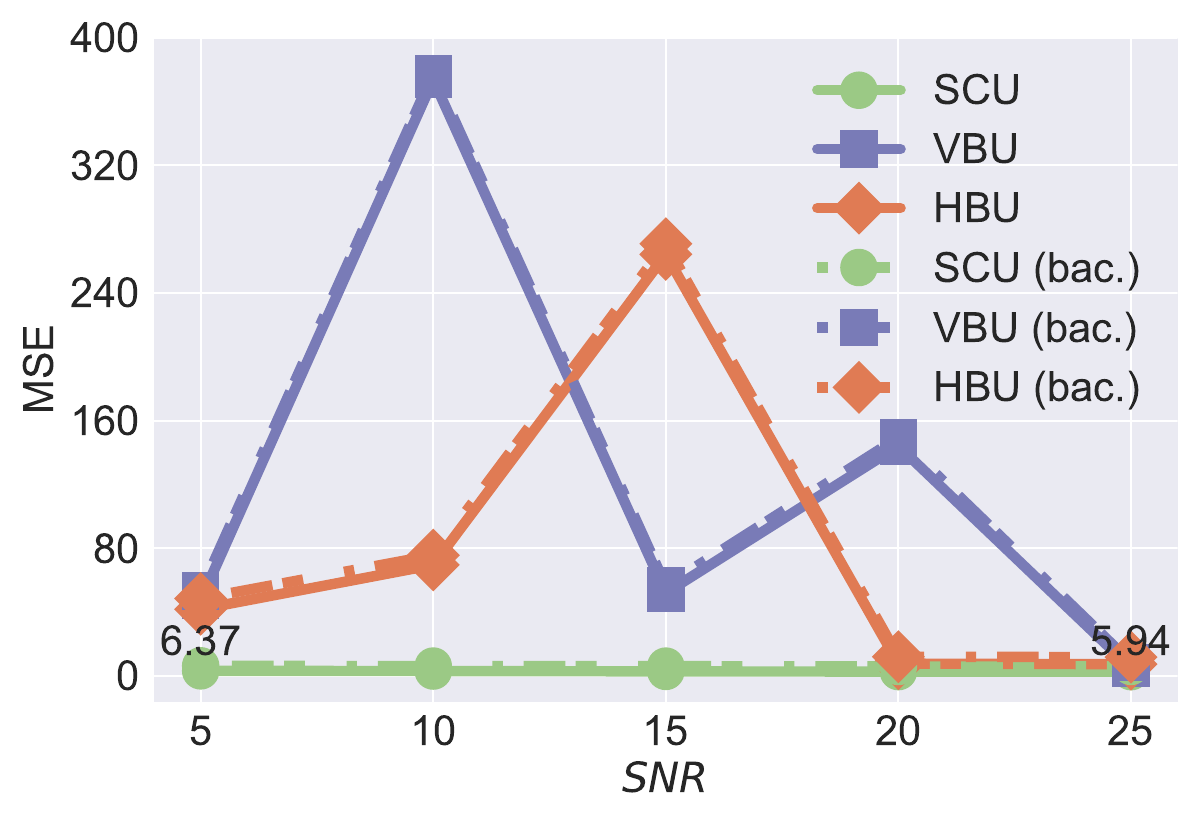}
	}
	\\
	\subfloat[\footnotesize Running time on MNIST, AWGN]{ \label{fig_mnistrtsnrawgnbar}
		\includegraphics[scale=0.28]{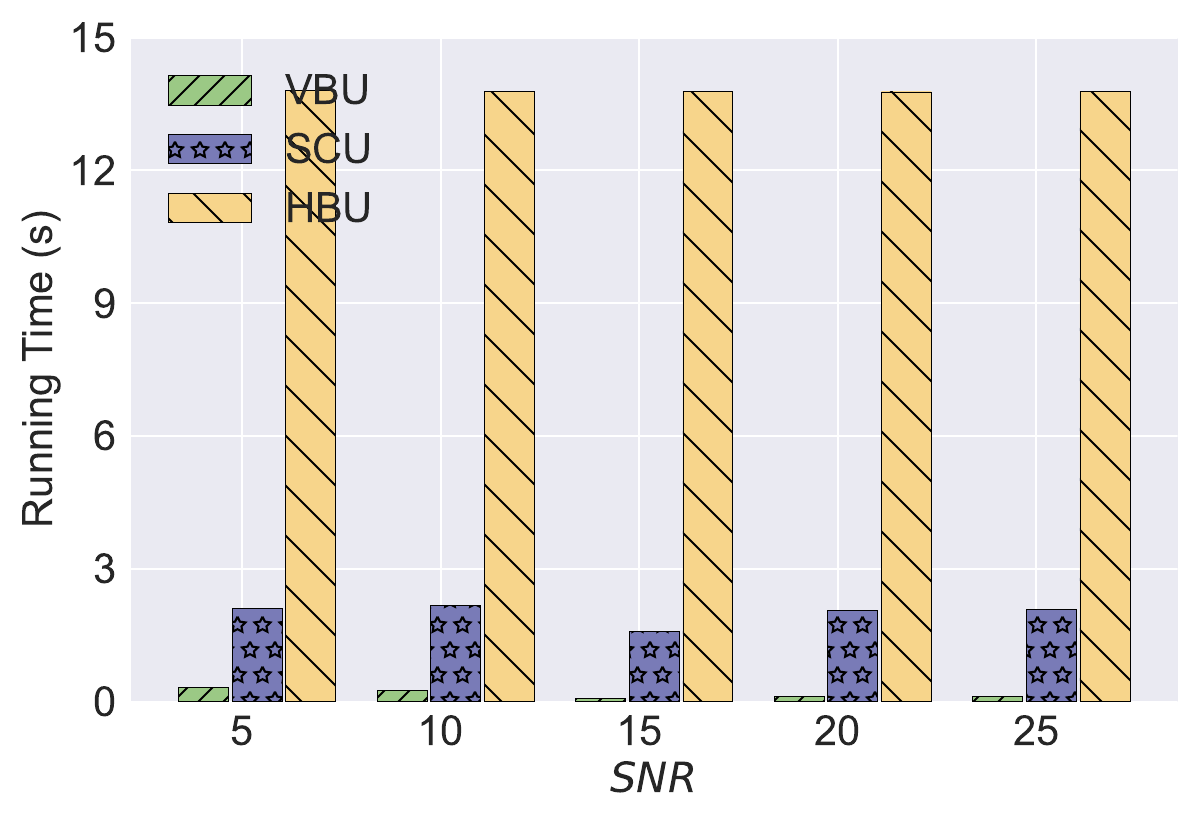}
	}
	\subfloat[\footnotesize Running time on MNIST, Rayleigh]{ \label{fig_mnistrtsnrrayleighbar}
		\includegraphics[scale=0.28]{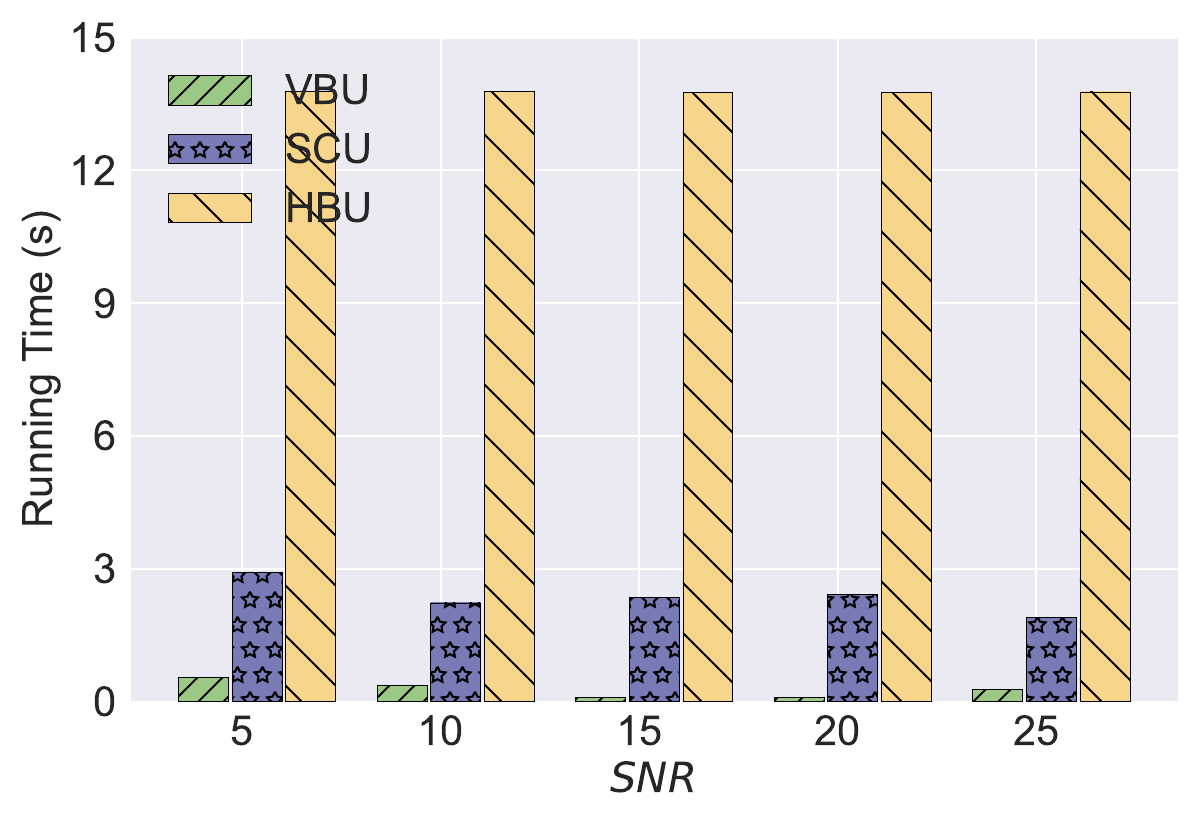}
	}
	\subfloat[\footnotesize Running time on MNIST, Rician]{ \label{fig_mnistrtsnrricianbar}
		\includegraphics[scale=0.28]{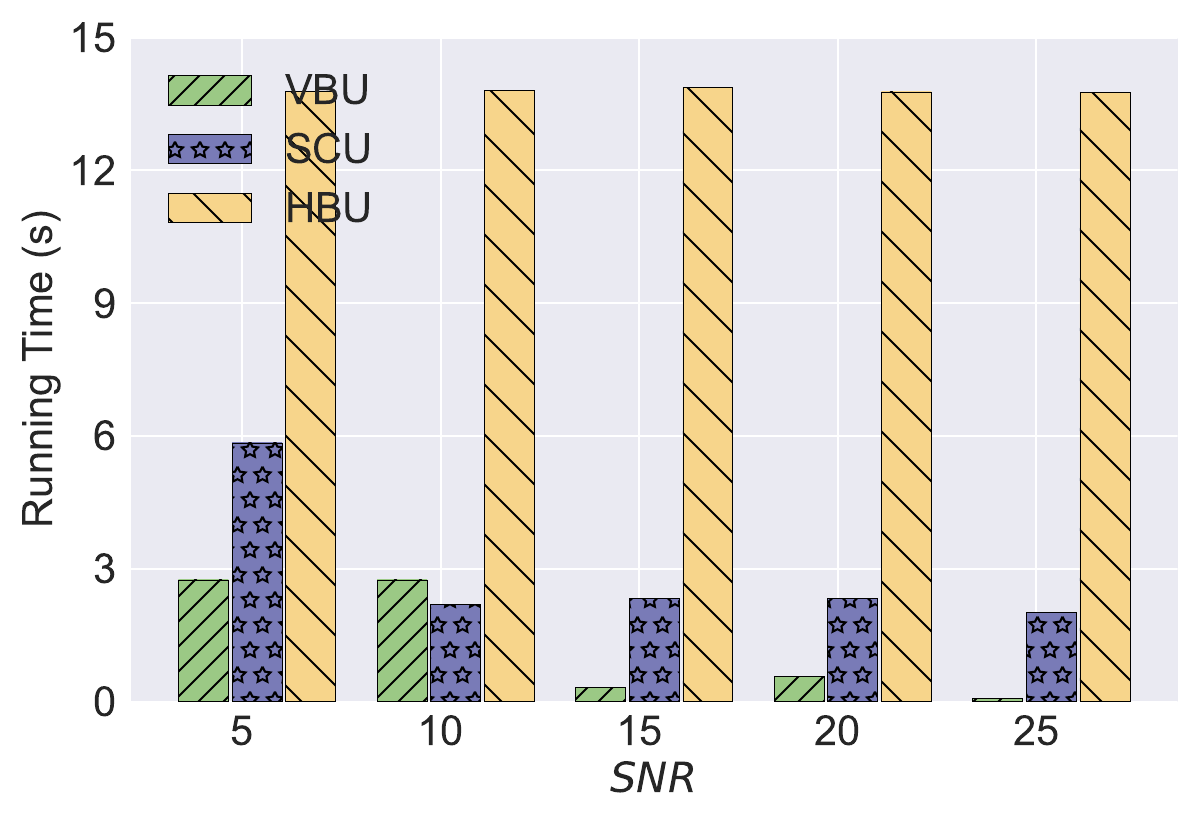}
	}
	\caption{Evaluations of the impact of different Channels and ${\it SNR}$. When ${\it SNR}$ increases, less noise is injected, and the unlearning effectiveness and model utility increases apparently.} 
	\label{mnist_evaluation_of_channel} 
\end{figure*}

\subsection{Evaluation Overview} \label{exp_overview}
We first present the general evaluation results of different methods on MNIST, CIFAR10 and CIFAR100, displayed in Table \ref{tab_total}.

\vspace{2mm}
\noindent
\textbf{Setup.} 
In these experiments, conducted across MNIST, CIFAR10, and CIFAR100 datasets, the erased data ratio (${\it EDR}$) is set at 6\% of the full original dataset, with the addition of AWGN channels. On MNIST, we set the ${\it SNR}=5$, while, on CIFAR10 and CIFAR100, we set the ${\it SNR}=20$. We set a higher ${\it SNR}$ for CIFAR10 and CIFAR100 because these two datasets are more complex than MNIST. More noise, such as ${\it SNR}=5$, decreases the semantic communication model utility worse. We assess the unlearning effect from four aspects: accuracy of the downstream model on the decoded clean data, backdoor accuracy on the decoded erased data and decoding MSE on the clean and the erased data. {The clean test dataset is the dataset prepared for testing and not in the training dataset. The backdoored erased dataset is in the original training dataset but needs to be unlearned later.} We evaluate the efficiency by the running time of the methods. 


\vspace{2mm}
\noindent
\textbf{Evaluation of Effectiveness.} 
In Table \ref{tab_total}, every method can effectively unlearn the trained semantic communication models. This is reflected in reduced backdoor accuracy and increased decoding MSE on the erased samples. All unlearning approaches effectively reduce the backdoor accuracy to below 10\%, a figure that is lower than what would be expected from random guessing. Although HBU and VBU achieve a better backdoor accuracy reduction on all datasets, they cause catastrophic unlearning, making the decoded samples useless for the downstream model. In evaluating the accuracy of the downstream model on decoded clean samples, baselines of both HBU and VBU result in an accuracy degradation of more than 20\% on these datasets.

Only SCU can mitigate the downstream model accuracy degradation on the decoded clean samples. 
On MNIST, SCE preserves the learned knowledge of the unlearned semantic models and enables the downstream model accuracy to be almost the same as the original performance. 
From the recovering aspect, all the decoding MSE on the erased samples are larger than that on the clean samples, which means all the unlearning methods have the unlearning effect. However, HBU and VBU increase the MSE too much on both erased and clean samples. It is the catastrophic unlearning of the semantic decoder, which means that these unlearning methods destroy the semantic decoding function of the decoder. And it is finally reflected in the huge accuracy degradation of the downstream models.


\vspace{2mm}
\noindent
\textbf{Evaluation of Efficiency.} 
The data presented in Table \ref{tab_total} reveal that every unlearning technique achieves a speed improvement greater than $10\times$ when compared to the time required to train the original model. It's noted that retraining from scratch consumes a similar amount of time as the original model training. HBU demonstrates inferior performance compared to the other two unlearning methods, attributable to its dependency on calculating the Hessian matrix using the remaining dataset. VBU emerges as the most efficient in terms of performance. Although SCU requires more running time than VBU, it is significantly faster than HBU across all datasets.


\begin{figure*}[t]
	\centering
	\subfloat[\footnotesize On MNIST]{ \label{fig_mnistepochmsetemp}
		\includegraphics[scale=0.32]{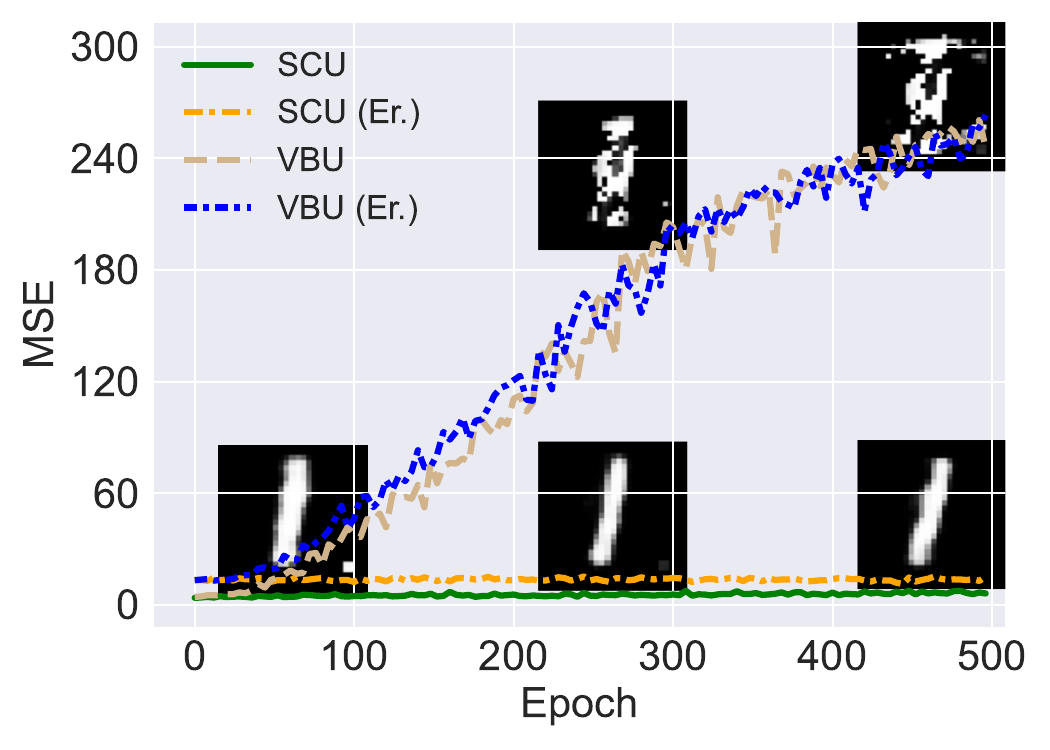}
	}
	\subfloat[\footnotesize On CIFAR10]{ \label{fig_cifarepochmsetemp}
		\includegraphics[scale=0.32]{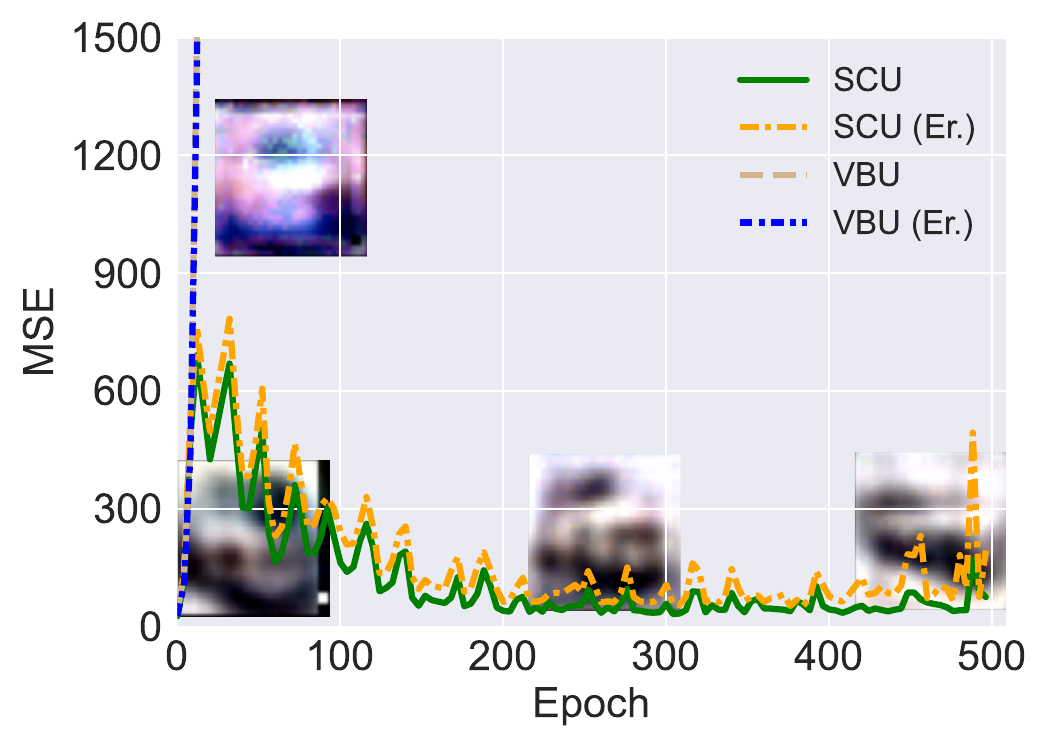}
	}
	\subfloat[\footnotesize On CIFAR100]{ \label{fig_cifar100epochmsetemp}
		\includegraphics[scale=0.32]{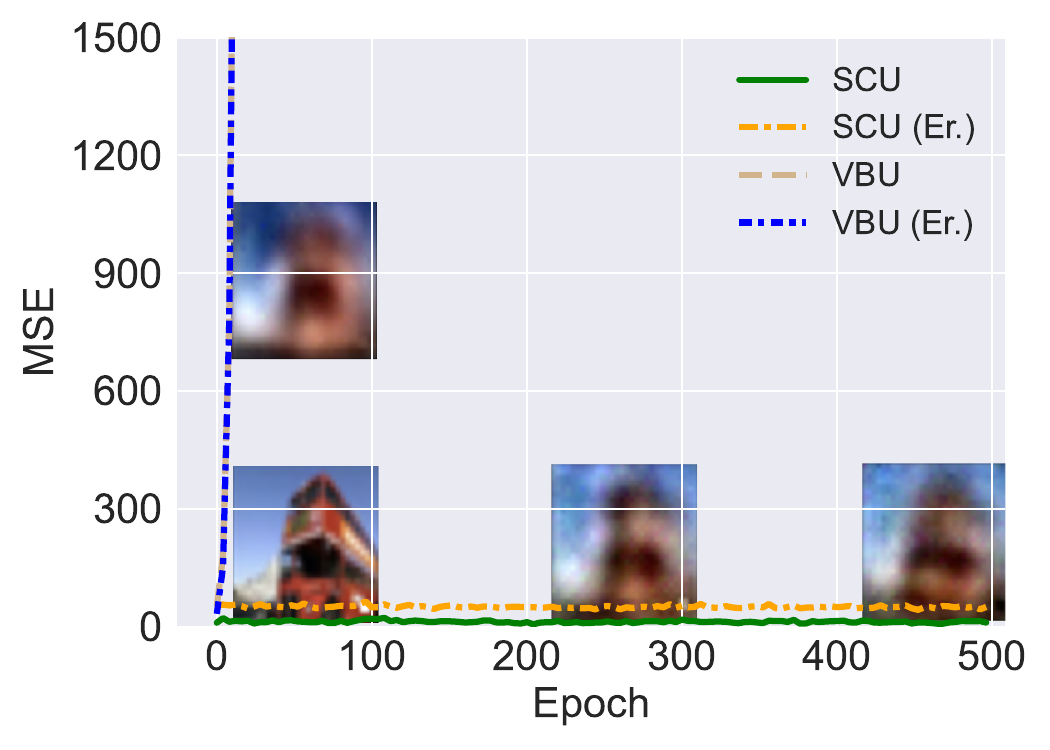}
	}
	\caption{Detailed changes during unlearning training processes and optimization results. On three datasets, SCU and SCU (Er.) keep a small MSE during the training process, indicating a good model utility preservation. The MSE of VBU and VBU (Er.) increases as the training continues. Especially on CIFAR10 and CIFAR100, the MSE of VBU and VBU (Er.) increases out of the scale of the vertical axis within 20 epochs.} 
	\label{detailed_changes_during_training} 
\end{figure*}


\subsection{Impact of the Erased Data Ratio} \label{exp_edr}

In this paper, three key variables – ${\it EDR}$ (erased data ratio), ${\it SNR}$ (signal-to-noise ratio), and the types of communication channels – significantly affect model performance. Our first evaluation focuses on the impact of ${\it EDR}$.

\vspace{2mm}
\noindent
\textbf{Setup.} 
With a fixed ${\it SNR}$ value and using AWGN channels, we explore the effects of varying ${\it EDR}$ values. The results of this exploration on the MNIST, CIFAR10, and CIFAR100 datasets are presented in Figure \ref{evaluation_of_edr}, {where ${\it EDR}$ is ranged from $2\%$ to $10\%$. In practice, we believe even $2\%$ data might be very large for unlearning. The results in \cite{bertram2019five} present that 3.2 million requests for deleting URLs have been issued to Google from 2014 to 2019, constituting less than $2\%$ of the total URLs Google indexes in the 5 years.} The evaluations of effectiveness are measured by the accuracy of the downstream model on the decoded clean samples in Figures \subref{fig_mnistaccercurve}, \subref{fig_cifaraccercurve} and \subref{fig_cifar100accercurve}, the backdoor accuracy of the downstream model on the decoded erased samples in Figures \subref{fig_mnistbackaccercurve}, \subref{fig_cifarbackaccercurve} and \subref{fig_cifar100backaccercurve}. We also demonstrate the decoding MSE on the clean and erased samples in Figures \subref{fig_mnistmsecleanercurve}, \subref{fig_mnistmseerasedercurve}, \subref{fig_cifarmsecleanercurve}, \subref{fig_cifarmseerasedercurve}, \subref{fig_cifar100msecleanercurve} and \subref{fig_cifar100mseerasedercurve}. The last column of Figure \ref{evaluation_of_edr} shows the running time of all methods on MNIST, CIFAR10, and CIFAR100.


\vspace{2mm}
\noindent
\textbf{Effectiveness of Methods.} 
We first analyze the unlearning methods' effectiveness by observing the accuracy of the downstream model on the decoded clean samples and backdoor accuracy on the decoded erased samples, which can directly demonstrate the backdoor removal effect and original semantic knowledge preservation of the unlearned semantic models. 
On all datasets, shown in Figures \subref{fig_mnistaccercurve}, \subref{fig_cifaraccercurve} and \subref{fig_cifar100accercurve}, as ${\it EDR}$ increases, the accuracy of almost every unlearned model decreases. 
Only SCU has a slight accuracy degradation, while HBU and VBU cause catastrophic unlearning when ${\it EDR}$ is large. 
From the backdoor erasure perspective, shown in Figures \subref{fig_mnistbackaccercurve}, \subref{fig_cifarbackaccercurve} and \subref{fig_cifar100backaccercurve}, all unlearning methods effectively eliminate the influence of the backdoor, reducing the backdoor accuracy to below 10\%, which is lower than what would be achieved by random selection. Although SCU is slightly higher than VBU and HBU on MNIST, it still decreases the backdoor accuracy from around $100\%$ to lower than $10\%$.

From the evaluation of decoding MSE, both HBU and VBU cause catastrophic unlearning, reflected in the much increased decoding MSE shown in Figure \subref{fig_mnistmsecleanercurve}, \subref{fig_mnistmseerasedercurve}, \subref{fig_cifarmsecleanercurve}, \subref{fig_cifarmseerasedercurve}, \subref{fig_cifar100msecleanercurve} and \subref{fig_cifar100mseerasedercurve}. {On CIFAR10 and CIFAR100, the decoding MSE of HBU increases too much, higher than the maximum value on the y-axis. We infer that the reason is HBU uses the Hessian matrix to calculate the data influence based on the remaining dataset but only updates the unlearned model in one step. It is easy to make the update overlarge, hence causing catastrophic unlearning. Most Hessian matrix-based unlearning methods need an update threshold to limit model updates and avoid utility degradation. In our experiments, we also set a fixed threshold for HBU. However, the fixed threshold is still easy to cause catastrophic unlearning.}
We show the decoding MSE of SCU on clean and erased samples in the third and fourth columns in \Cref{evaluation_of_edr}. The MSE on erased samples is all higher than that on clean samples. It proves that SCU achieves unlearning effectiveness from another aspect besides the backdoor accuracy.

\vspace{2mm}
\noindent
\textbf{Efficiency of Methods.} 
The efficiency of each method is assessed based on the running times depicted in Figures \subref{fig_mnistrterbar}, \subref{fig_cifarrterbar}, and \subref{fig_cifar100rterbar}. We observe a slight increase in running time with larger ${\it EDR}$ values, particularly noticeable in the MNIST dataset.
HBU requires the most time for unlearning, exceeding the duration needed by the other two methods by more than $5\times$. This increased time consumption is attributed to HBU's requirement for Hessian matrix calculation based on the remaining dataset.
In contrast, VBU is the most efficient in terms of reducing running time. SCU, while faster than HBU, takes about twice as long as VBU. This is because SCU incorporates an additional step of optimizing the contrastive compensation loss, which is based on the selected remaining dataset. However, it is still much faster than HFU. Moreover, compared with retraining and original training, all the methods achieve a significant speedup.

\subsection{Impact of the Channels and SNR} \label{exp_channels}

During semantic communication model training, the channels and Signal-to-Noise Ratio (${\it SNR}$) have an important influence on the model's training and performance.
These two will also have a huge impact on the training of unlearning methods for semantic communication models. Since the noise of different ${\it SNR}$ is also highly related to the channels, we demonstrate the results of different ${\it SNR}$ in different channels in the same section. We choose the AWGN, Rayleigh and Rician channels, and set the ${\it SNR}$ from 5 to 25, and all the results on MNIST are shown in Figure \ref{mnist_evaluation_of_channel}. 

\vspace{2mm}
\noindent
\textbf{Effectiveness impacted by ${\it SNR}$.} 
We demonstrate the downstream model accuracy on the clean samples and backdoor accuracy on the removed samples together in Figures \subref{fig_mnistaccsnrawgncurve}, \subref{fig_mnistaccsnrrayleighcurve} and \subref{fig_mnistaccsnrriciancurve}. As we can see, no matter in which channel, when ${\it SNR}$ is small, HBU and VBU have a huge utility degradation. Especially in Rician channels, when ${\it SNR} \leq 15$, the accuracy of HBU and VBU decreases to around 0, almost unusable. Only SCU achieves a satisfied performance with whatever ${\it SNR}$. It preserves the downstream model accuracy of more than $95\%$ and achieves a backdoor removal, ensuring backdoor accuracy of less than $10\%$. 

From the decoding MSE perspective, illustrated in Figures \subref{fig_mnistmsecleansnrawgncurve}, \subref{fig_mnistmsecleansnrrayleighcurve} and \subref{fig_mnistmsecleansnrriciancurve}, although all the unlearning methods increase the MSE on erased samples than on the clean samples, i.e., forgetting more information about the erased samples, HBU and VBU cause too much MSE increase. Especially when ${\it SNR}$ is small, both HBU and VBU cause a vast MSE increase on both clean and erased samples and make the decoded samples unusable in the downstream models. Compared with them, SCU only has a slight MSE increase when ${\it SNR}$ is small; we show the value of SCU (bac.) when ${\it SNR}=5$ and ${\it SNR}=25$ in Figures \subref{fig_mnistmsecleansnrawgncurve}, \subref{fig_mnistmsecleansnrrayleighcurve} and \subref{fig_mnistmsecleansnrriciancurve}. It effectively forgets the data intended for erasure while retaining the insights gained from the remaining samples, as evidenced by a significantly reduced MSE when compared to both VBU and HBU. 

\vspace{2mm}
\noindent
\textbf{Efficieny impacted by ${\it SNR}$.} 
From the running time perspective, shown in Figures \subref{fig_mnistrtsnrawgnbar}, \subref{fig_mnistrtsnrrayleighbar} and \subref{fig_mnistrtsnrricianbar}, when ${\it SNR}$ increases, the unlearning training usually consumes lesser time. It is most obvious in Rician channels. VBU achieves the best speedup and HBU consumes the most training time. Although SCU consumes time around triple than VBU, it also achieves more $3\times$ speedup than HBU. 

\subsection{Detailed Analysis during Training Processes} \label{exp_opt}

In this section, we present a detailed analysis of the training alterations that occur during the optimization of the proposed methods. This analysis aims to elucidate why SCU is able to outperform other state-of-the-art methods. We demonstrate the unlearning training process of SCU and VBU on MNIST, CIFAR10 and CIFAR100 in Figure \ref{detailed_changes_during_training}. The results include the decoding MSE of each batch training on clean and erased (abbreviated as Er.) samples. The results include the decoding MSE of each training batch on clean and erased (abbreviated as Er.) samples. VBU will easily cause catastrophic unlearning as the unlearning training continues, where the decoding MSE of VBU increases from around 2 to around 250 on MNIST and quickly increases more than the maximum of the vertical axis on CIFAR10 and CIFAR100. In contrast, SCU mitigates utility degradation and preserves the decoding of MSE at a satisfied value. Moreover, it keeps the decoding MSE on erased samples higher than that on clean samples, ensuring the forgetting effectiveness.


\subsection{Ablation Study}
\begin{figure}[t]
	\centering
	\hspace{-4mm}
	\subfloat[\footnotesize Accuracy ]{ 	\label{fig:mnistaccercurveablation}
		\includegraphics[scale=0.17]{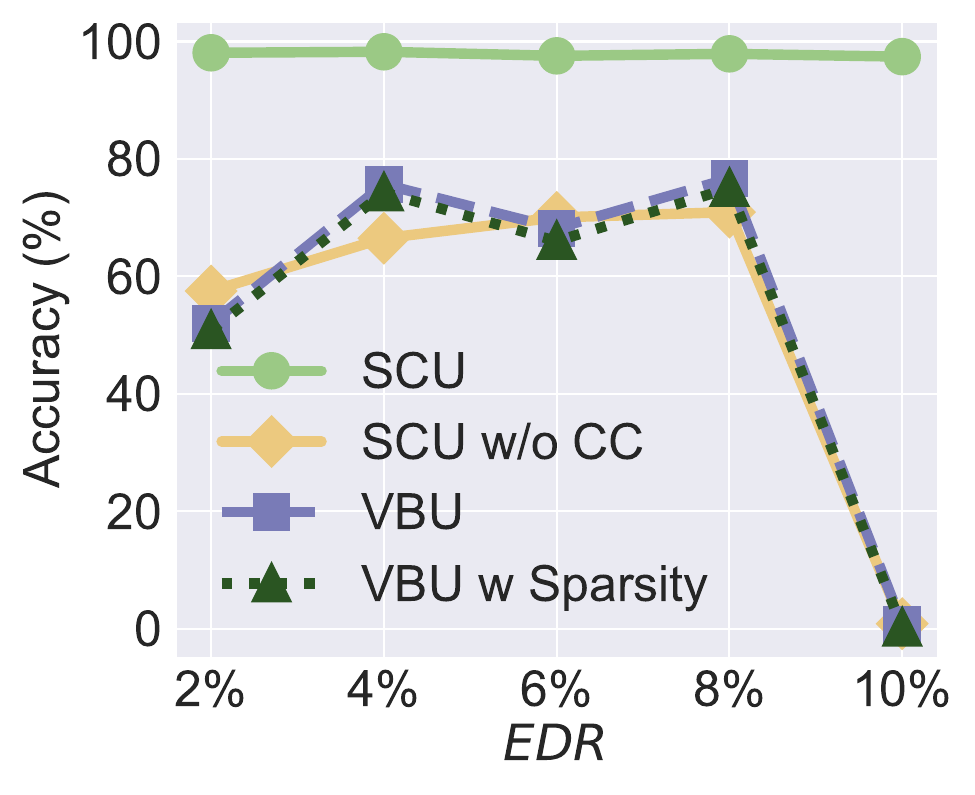}
	}	\hspace{-5mm}
	\subfloat[\footnotesize Bac. Acc.]{ 	\label{fig:mnistbackaccercurveablation} 
		\includegraphics[scale=0.17]{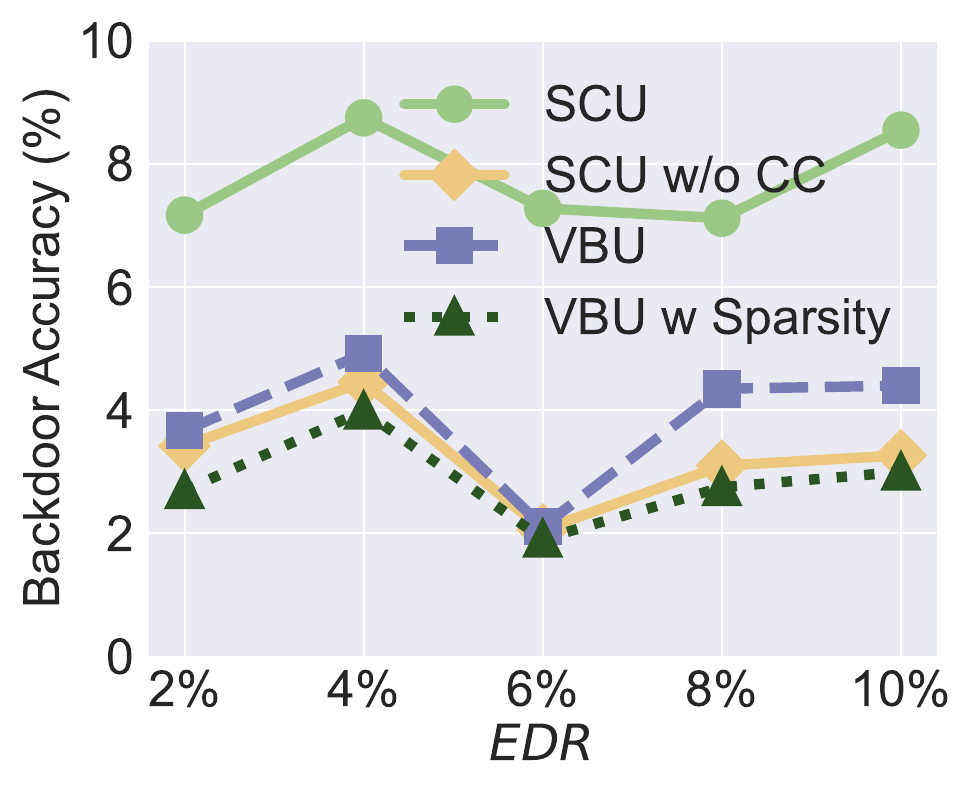}
	}	\hspace{-5mm}
	\subfloat[\footnotesize MSE]{	\label{fig:mnistmsecleanercurveablation}
		\includegraphics[scale=0.17]{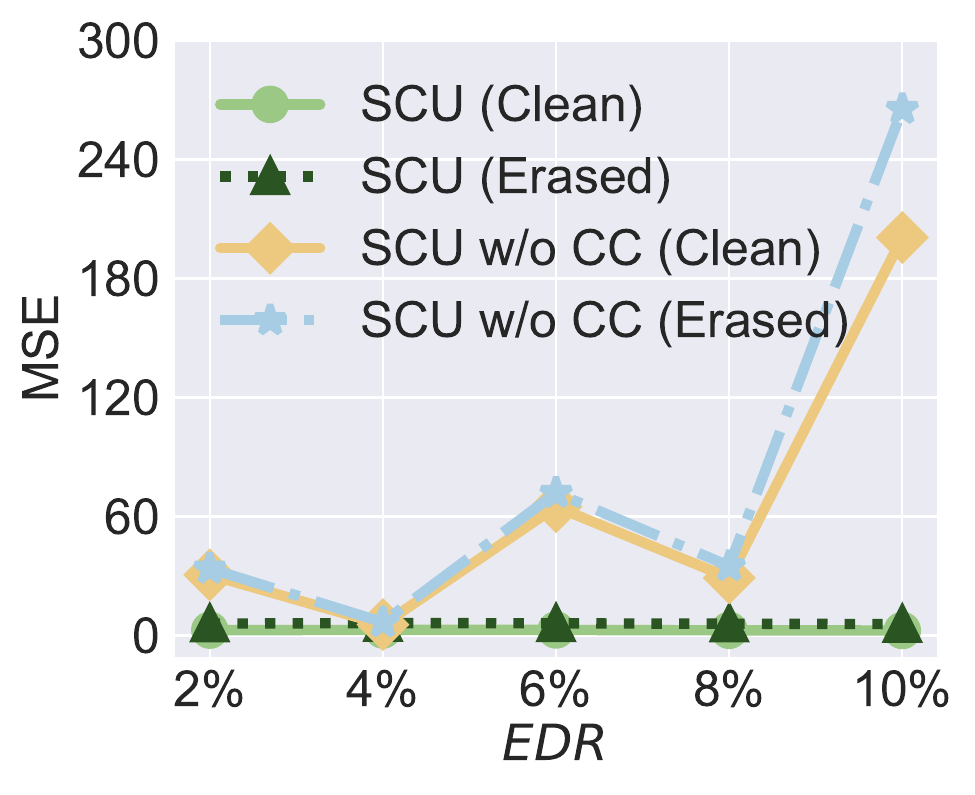}
	}
	\caption{Ablation study when SCU without (abbreviated as w/o) the CC loss on MNIST. It performs similarly to VBU, much worse than the entire SCU.} 
	\label{ablation_study} 
\end{figure}

We evaluate the impact of the CC loss on SCU. Specifically, we compare the performance of SCU and the performance of the SCU without the CC loss, denoted as ``SCU w/o CC''. {For comparison, we also conduct experiments on VBU and an extended VBU that considers the model sparsity as \cite{liu2024model}, denoted as ``VBU w Sparsity''.} We present the experimental results on MNIST in \Cref{ablation_study}. 

In Figures \subref{fig:mnistaccercurveablation} and \subref{fig:mnistbackaccercurveablation}, we can see that the ``SCU w/o CC'' performs similarly to VBU. {Without the contrastive compensation, the only joint unlearning is similar to VBU. Since both VIB and VAE have the same expansion mathematic form for optimization, the corresponding unlearning methods perform similarly without the CC compensation. The results of ``VBU w Sparsity'' show that the model sparsity does improve the unlearning effectiveness but has a slight model utility degradation. Compared with considering model sparsity, our method with contrastive compensation, SCU,  significantly preserves the model utility while slightly mitigating the unlearning effect.} And all three subfigures in \Cref{ablation_study} demonstrate that the improvement of SCU with the constructive compensation.

\section{Summary and Future Work} \label{summary}


This study explores the challenge of unlearning in semantic communication and presents the SCU method as a novel solution. 
Our approach contains two key steps. In the first step, we develop a joint unlearning strategy that simultaneously unlearns both the semantic encoder and decoder. To counteract the potential reduction in utility resulting from the unlearning process, we subsequently introduce a contrastive compensation method in the second step. The contrastive compensation method treats the erased data as the negative samples and the remaining data as the positive samples to execute contrastive retraining. Through rigorous analysis and comprehensive experiments, we assess the effectiveness and efficiency of the SCU method. Our findings confirm its superiority over existing state-of-the-art unlearning approaches used in semantic communication in semantic communication. 

In the future, although our SCU can effectively mitigate the negative impact of unlearning than existing methods, it still influences the downstream models' performance, especially in complex datasets CIFAR10 and CIFAR100. We will put more effort into investigating and trying to solve this problem. Moreover, there is still a largely uncharted area to be explored for designing unlearning methods for unsupervised and task-oriented DL models in semantic communications.

\footnotesize
\bibliographystyle{IEEEtranN}
\bibliography{semantic_c_unl}

\begin{IEEEbiography}[{\includegraphics[width=1in,height=1.25in,clip,keepaspectratio]{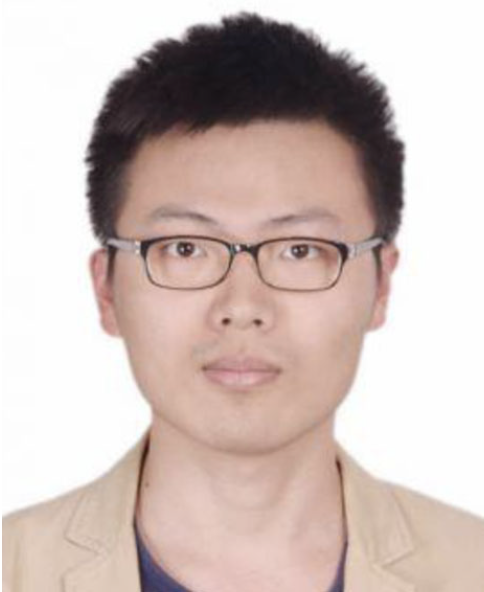}}]{Weiqi Wang}(IEEE M'24)
received the M.Sc. degree in computer science from Soochow University, Suzhou, China, in 2018. He is currently pursuing the Ph.D. degree with the School of Computer Science, University of Technology Sydney, where he is advised by Prof. Shui Yu. He previously worked as a senior algorithm engineer at the Department of AI-Strategy, Local consumer services segment, Alibaba Group. He has been actively involved in the research community by serving as a reviewer for prestige journals such as ACM Computing Surveys, IEEE Communications Surveys and Tutorials, IEEE TIFS, IEEE TDSC, IEEE TIP, IEEE Transactions on SMC, and IEEE IOTJ, and international conferences such as The ACM Web Conference (WWW), ICLR, IEEE ICC, and IEEE GLOBECOM. His research interests are in machine unlearning, federated learning, and security and privacy in machine learning.
\end{IEEEbiography}

\begin{IEEEbiography}[{\includegraphics[width=1in,height=1.25in,clip,keepaspectratio]{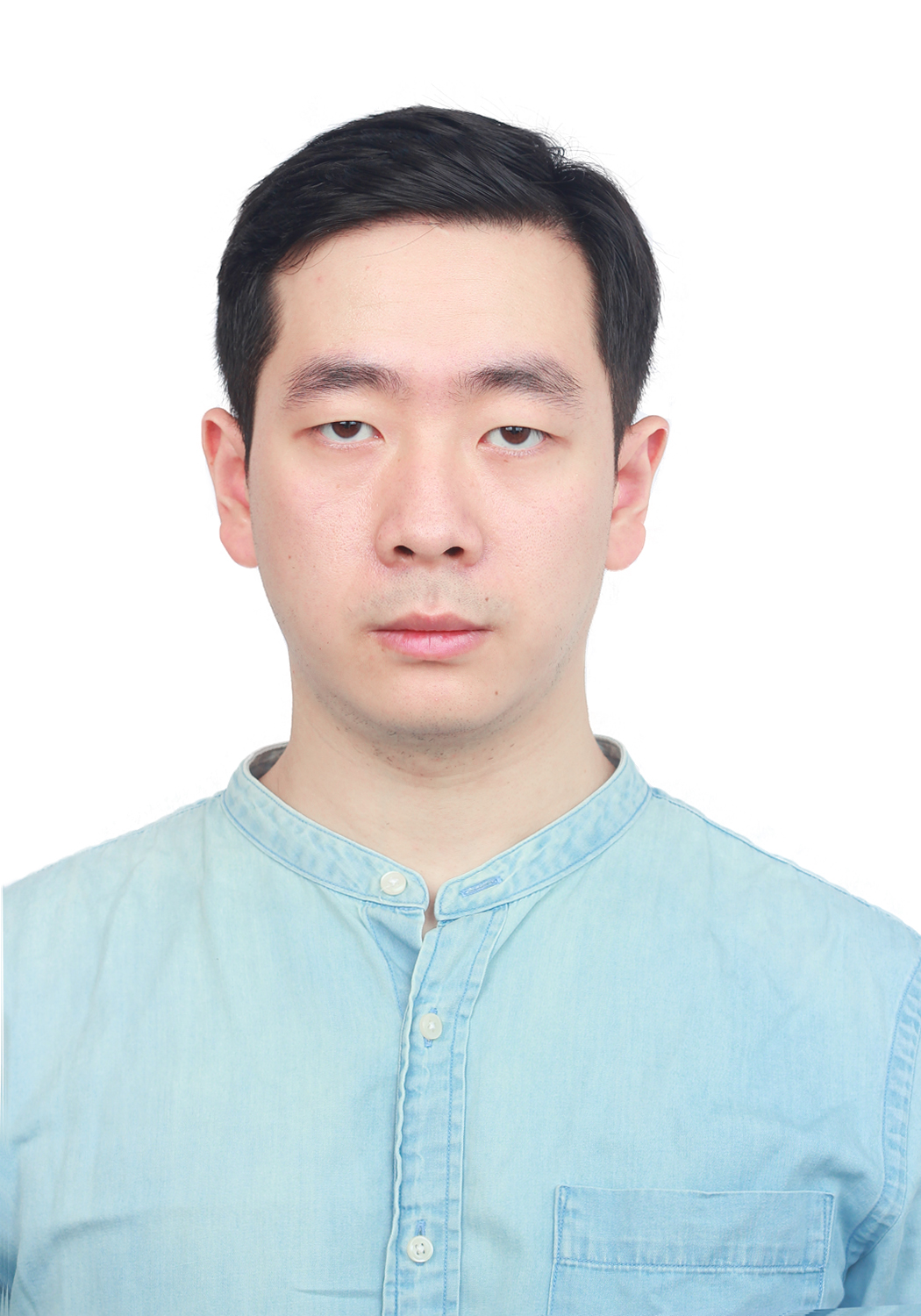}}]{Zhiyi Tian} (IEEE M'24) received the B.S. degree and the M.S. degree from Sichuan University, China, in 2017 and 2020, respectively. He received the Ph.D. degree in 2024 from University of Technology Sydney, Australia. He currently is a research associate of University of Technology Sydney, Australia. His research interests include security and privacy in deep learning, semantic communications. He has been actively involved in the research community by serving as a reviewer for prestige journals, such as ACM Computing Surveys, IEEE Communications Surveys and Tutorials, TIFS, TKDD, and international conferences, such as IEEE ICC and IEEE GLOBECOM.
\end{IEEEbiography}

\begin{IEEEbiography}[{\includegraphics[width=1in,height=1.25in,clip,keepaspectratio]{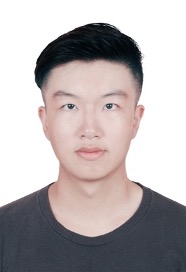}}]{Chenhan Zhang} (IEEE S'19 M'24) obtained his Ph.D. from University of Technology Sydney, Australia, in 2024, where he was advised by Prof. Shui Yu. Before that, he obtained his B.Eng. (Honours) from University of Wollongong, Australia, and  M.S. from City University of Hong Kong, Hong Kong, in 2017 and 2019, respectively. He is currently a postdoctoral research fellow at Cyber Security Hub, Macquarie University, Australia. His research interests include security and privacy in graph neural networks and trustworthy spatiotemporal cyber physical systems. He has been actively involved in the research community by serving as a reviewer for prestige venues such as ICLR, IJCAI, INFOCOM, IEEE TDSC, IEEE IoTJ, ACM Computing Survey, and IEEE Communications Surveys and Tutorials.
\end{IEEEbiography}
\begin{IEEEbiography}[{\includegraphics[width=1in,height=1.25in,clip,keepaspectratio]{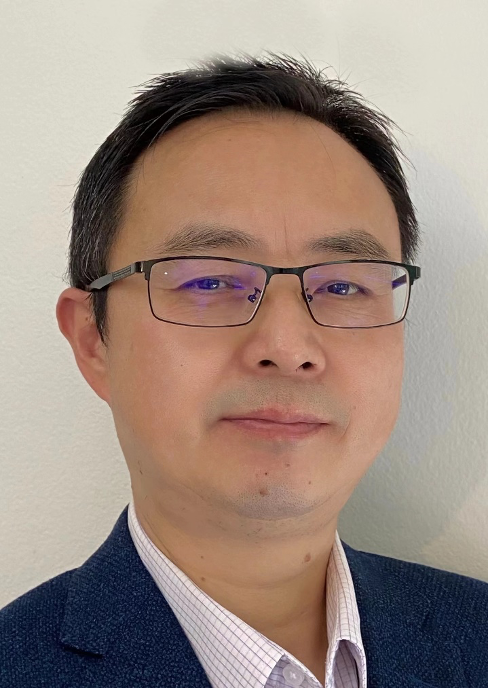}}]{Shui Yu} (IEEE F’23) obtained his PhD from Deakin University, Australia, in 2004. He is a Professor of School of Computer Science, Deputy Chair of University Research Committee, University of Technology Sydney, Australia. His research interest includes Cybersecurity, Network Science, Big Data, and Mathematical Modelling. He has published five monographs and edited two books, more than 600 technical papers at different venues, such as IEEE TDSC, TPDS, TC, TIFS, TMC, TKDE, TETC, ToN, and INFOCOM. His current h-index is 80. Professor Yu promoted the research field of networking for big data since 2013, and his research outputs have been widely adopted by industrial systems, such as Amazon cloud security. He is currently serving the editorial boards of IEEE Communications Surveys and Tutorials (Area Editor) and IEEE Internet of Things Journal (Editor). He served as a Distinguished Lecturer of IEEE Communications Society (2018-2021). He is a Distinguished Visitor of IEEE Computer Society, and an elected member of Board of Governors of IEEE VTS and ComSoc, respectively. He is a member of ACM and AAAS, and a Fellow of IEEE. 
\end{IEEEbiography}

\vfill

\end{document}